\documentclass[prb,twocolumn,showpacs]{revtex4}
\usepackage{graphicx}
\usepackage{dcolumn}
\usepackage{bm}
\usepackage{amsmath}

\begin{document}

\title{Multi-reference symmetry-projected variational approximation for
the ground state of the doped one-dimensional Hubbard model}

\author{R. Rodr\'{\i}guez-Guzm\'an$^{1,2}$, Carlos A. Jim\'enez-Hoyos$^{1}$ and Gustavo E. Scuseria$^{1,2}$}

\affiliation{
$^{1}$ Department of Chemistry, Rice University, Houston, Texas 77005, USA
\\
$^{2}$ Department of  Physics and Astronomy, Rice University, Houston, Texas 77005, USA
}

\date{\today}

\begin{abstract}
A multi-reference configuration mixing scheme is used to describe the
ground state, characterized by well defined spin and space group symmetry quantum
numbers as well as doping fractions $N_{e}/N_{sites}$, of one dimensional   Hubbard lattices with
nearest-neighbor hopping and periodic boundary conditions.  Within this scheme, each ground state is expanded
in a given number of nonorthogonal and  variationally determined
symmetry-projected configurations. The
results obtained  for  the ground state and correlation energies of
half-filled and doped lattices  with 30, 34 and 50 sites, compare well with
 the exact
Lieb-Wu solutions as well as with  the ones obtained with other state-of-the-art approximations.
The structure of the intrinsic
symmetry-broken determinants resulting from the variational procedure is interpreted
in terms of solitons whose translational and breathing motions can be regarded as
 basic units of quantum fluctuations. It is also shown that in the case of doped 1D lattices,
 a part of such fluctuations can also be interpreted in terms of polarons.
In addition to momentum distributions, both spin-spin and density-density
correlation
functions are
studied as functions of doping. The spectral functions and density of states,
computed with an ansatz whose quality can be well-controlled by the number
of symmetry-projected configurations used to approximate the
$N_{e} \pm 1$ electron systems, display features beyond a simple quasiparticle
distribution, as well as spin-charge separation trends.
\end{abstract}


\maketitle

\section{Introduction.}
\label{INTRO}

Disentangling the effects of electron-electron
interactions in the ground and excited states of
low-dimensional  systems has become
an exciting challenge in contemporary condensed matter physics. \cite{Dagotto-review,Dagotto-Rev-Mod-Physics-2013}
In particular, the discovery of high-Tc superconductivity
in the cuprates
\cite{HTCSC-1}
has acted as a driving force
to develop  theoretical  models able to account for the most
relevant correlations in many-electron systems in the simplest possible
way. Within this context, the  repulsive
Hubbard model \cite{Hubbard-model_def1} has been widely studied
for several reasons including that it represents a prototype
for the Mott transition between a metal and an antiferromagnetic
insulator \cite{Vollhardt} and the suggestion  \cite{Anderson-1} that it contains
the
basic physics associated with high-Tc superconductivity.
The phenomenon of colossal magnetic resistance \cite{Science-Dagotto}
has also attracted considerable attention. On the other hand, the
study of the  high-Tc iron-based superconductors
\cite{sup-Fe,Stewart-Review}
has become a very active research area. \cite{Su-Fe-Dagotto}  Here, calculations
in terms of multi-orbital
Hubbard-like
models have already provided  valuable insight into the
interplay between  doping and the strength of the electronic
correlations in these exotic superconductors. \cite{Dagotto-Rev-Mod-Physics-2013}
Hubbard models represent valuable tools
to study cold fermionic atoms
in optical lattices \cite{optical-1} as well as the properties of
graphene. \cite{CastroNeto-review} It has also become
clear that their strong coupling limits, i.e., the Heisenberg models, \cite{text-Hubbard-1D} can
be quite useful
to study low-dimensional magnets
whose properties might be relevant for real materials found
in nature and/or synthesized by means of crystal growing. \cite{Mikeska}

The previous examples already show the
central role of
Hubbard-like lattice models and their strong
coupling versions to obtain insight into the
properties associated with the emergent complexity
in many-electron problems. Precisely, it is this complexity
that requires the
use of different approximations to describe
low-dimensional systems. Among  the available theoretical tools we have, for example,
exact diagonalizations for small
lattices \cite{Dagotto-review,Lanczos-Fano} while for the
larger ones, we  can resort to
quantum Monte Carlo, \cite{Nightingale,Raedt-MC} variational
Monte Carlo, \cite{Neuscamman-2012} coupled cluster, \cite{Bishop-1,Bishop-2}
variational reduced-density-matrix \cite{var-red-den-mat} and
density matrix renormalization
group \cite{DMRG-White,Dukelsky-Pittel-RPP,Scholl-RMP}
methods,
as well as approximations based on
matrix product
and tensor network states. \cite{Scholl-AP,GChan,TNPS-1,TNPS-2} Both
frequency-dependent and frequency-independent embedding approaches
\cite{Zgid,DMFT-1,maier2005,stanescu2006,moukouri2001,huscroft2001,aryanpour2003,DVP-2,Knizia-Chan,Irek-DMET}
are also actively pursued.

\begin{table*}
\label{Table1}
\caption{ Ground state energies for
half-filled and doped lattices with $N_{sites}$=30 sites
and $N_{e}$=14, 18, 22, 26, 30 electrons predicted with the
FED scheme  ($n$=60  GHF transformations for $U$=2$t$, 4$t$ and $n$=150 for $U$=8$t$)
are compared with exact values.
Results obtained with
the FED$^{*}$  [$n$=25 GHF transformations] (Ref. 51),
and
ResHF  [$n$=30 UHF transformations] (Ref. 41) approximations
as well as
the
energies corresponding to  the RHF and
other HF  solutions  are also included in the table.
The ratio of correlation energies $\kappa$, is computed according to Eq.(\ref{formulaCE}).
}
\begin{tabular}{cccccccccccccccccccc}
\hline
\\
$U$    & &   &        & & $N_{e}$=30  &$\kappa(\%)$  & &  $N_{e}$=26	&$\kappa(\%)$	&   &	   & $N_{e}$=22   &$\kappa(\%)$ &    & $N_{e}$=18  &$\kappa(\%)$ &    &  $N_{e}$=14 &$\kappa(\%)$  \\
\\
\hline
\\

 2$t$    & &   & RHF       & & -23.2671    &  	       &&  -26.1642      &	       &    &	   & -26.8921	  &		&    & -25.5587    &		 &    & -22.3390    &		   \\
       & &   & HF        & & -23.4792    & 10.02       &&  -26.1642      &    0	       &    &	   & -26.8921	  &  0  	&    & -25.5587    &  0 	 &    & -22.3390    &  0      \\		
       & &   & FED       & & -25.3800    & 99.83       &&  -28.0201      &   99.72     &    &	   & -28.4391	  & 99.68	&    & -26.7816    & 99.69	 &    & -23.2365    & 99.74	      \\
       & &   &FED$^{*}$  & & -25.3730    & 99.50       &&  -             &   -         &    &      &  -           & -           &    &  -          & -           &    &  -          &     \\
       & &   & ResHF     & & -25.3436    & 98.11       &&  -27.9979      &   98.52     &    &      & -28.4268     & 98.88       &    &  -          &  -          &    &  -          & -      \\         				
       & &   & Exact     & & -25.3835    &  	       &&  -28.0253      &	       &    &	   & -28.4441	  &		&    & -26.7854    &		 &    & -23.2388    &	       \\
\\
\hline
\\
 4$t$     & &  & RHF       &  & -8.2671    &	       && -14.8975	&		&    &     & -18.8254	  &		&    & -20.1587    &		 &    &  -19.0723   &			\\
        & &  & HF        &  & -14.0732   & 64.75       && -17.3756	&  34.96	&    &     & -20.1328	  & 23.42	&    & -21.1011    & 22.45	 &    &  -19.8855   &  27.84	   \\			
        & &  & FED       &  & -17.2081   & 99.71       && -21.9193	&  99.04	&    &     & -24.3497	  & 99.03	&    & -24.3222    & 99.19	 &    &  -21.9824   &  99.63	  \\
        & &  &FED$^{*}$  &  & -17.1789   & 99.39       &&  -            &   -           &    &     &  -           & -           &    &  -          & -           &    &  -          &     \\  	
	& &  & ReSHF     &  & -17.0542   & 98.00       && -21.5720      &  94.15        &    &     & -24.1582     & 95.56       &    &  -          &   -         &    &   -         & -        \\ 	   				
        & &  & Exact     &  & -17.2335   &	       && -21.9868	&		&    &     & -24.4057	  &		&    & -24.3561    &		 &    &  -21.9932   &			\\
\\
\hline
\\
 8$t$     & &  & RHF       &  &  21.7329   &	       &&  7.6358 	&		&    &      & -2.6921	   &		&    &  -9.3587    &		 &    & -12.5390    &		\\
        & &  & HF        &  & -7.8329    & 93.65       && -11.2049	& 79.45 	&    &      & -14.9299     & 69.32	&    & -18.1922    & 70.12	 &    & -19.0005    &  77.96	    \\
        & &  & FED       &  & -9.8260    & 99.95       && -15.8927	& 99.22 	&    &      & -20.1711     & 99.01	&    & -21.8777    & 99.38	 &    & -20.8062    &  99.75	       \\
        & &  &FED$^{*}$  &  & -9.7612    & 99.75       &&  -            &   -           &    &      &  -           & -          &    &  -          & -           &    &  -          &     \\  		
	& &  & ResHF     &  & -9.5378    & 98.46       && -15.4059      & 97.17         &    &      & -19.5552     & 95.52      &    &  -          &  -          &    &  -          & -      \\	  		
        & &  & Exact     &  & -9.8387    &	       && -16.0761	&		&    &      & -20.3462     &		&    & -21.9555    &		 &    & -20.8271    &		     \\
\\
\hline
\end{tabular}
\end{table*}

The exact Bethe ansatz solution of the one-dimensional (1D) Hubbard Hamiltonian is well
known, \cite{LIEB,BETHE} which is not the case for the two-dimensional (2D) model.
On the other hand, in spite of the considerable progress already made, the exact 1D wave
 functions still remain difficult to handle in practice when computing several physical
properties. \cite{text-Hubbard-1D} It has also remained difficult  to obtain
an intuitive physical picture of the basic units of quantum fluctuations
\cite{Tomita-3,Tomita-1,Tomita-2,Tomita-2011PRB}
using the  Lieb-Wu solutions \cite{LIEB}  and/or within the theoretical
frameworks already mentioned above. This task is further
complicated by the fact that quantum fluctuations
can exhibit
unconventional features in low-dimensional systems. A typical example, is
the spin-charge separation in the strong coupling regime
 of the 1D Hubbard
model. \cite{text-Hubbard-1D,Ogata-Shiba,Voit} Angle-resolved photoemission
spectroscopy  results
also reveal a complex pattern of spin-charge coupling/decoupling  both in 1D and 2D
systems \cite{Kim,KMShen} in the weak and intermediate-to-strong interaction
regimes.

Therefore, it is highly desirable to explore
the performance of
alternative wave function based
approaches that, on the one hand, could complement
existing state-of-the-art theoretical tools and, on the other hand, lead us to compact states
whose (intrinsic) structures
are simple enough
to be interpreted in terms of basic units of quantum fluctuations in
half-filled and doped lattices. Within this context, the 1D Hubbard model
represents a challenging testing ground since both its exact solution and
highly accurate density matrix renormalization
group results are available.
The last years have seen some progress
in this direction within the framework of the single-reference
(SR) and multi-reference (MR) symmetry-projected approximations
\cite{Tomita-1,Tomita-2,Tomita-2011PRB,Carlos-Hubbard-1D,Rayner-2D-Hubbard-PRB-2012,non-unitary-paper-Carlos,rayner-Hubbard-1D-FED2013}
which are routinely used in nuclear structure physics. \cite{rs,Carlo-review,rayner-GCM-paper,rayner-GCM-parity}
Note that in quantum chemistry the names single-component and multi-component have been adopted 
\cite{Carlos-Rayner-Gustavo-VAMPIR-molecules,Carlos-Rayner-Gustavo-FED-molecules} 
instead of single-reference and multi-reference , respectively.

Recently,  a hierarchy of symmetry-projected
variational approaches  \cite{Carlo-review}
has been applied
to describe both the ground and excited states
of the 1D and 2D Hubbard models. \cite{Carlos-Hubbard-1D,Rayner-2D-Hubbard-PRB-2012,non-unitary-paper-Carlos,rayner-Hubbard-1D-FED2013}
 In its simplest (i.e., SR) form, the symmetry-projected variation-after-projection (VAP)
 method \cite{Carlos-Hubbard-1D}
resorts to a Hartree-Fock-type \cite{rs} (HF)  trial state  $| {\cal{D}} \rangle$ that
deliberately breaks  several symmetries of the considered
Hamiltonian. The method then  superposes, with the
help of projection operators, \cite{Carlos-Hubbard-1D} a degenerate manifold
of Goldstone states
$\hat{R}| {\cal{D}} \rangle$, with $\hat{R}$ being a
symmetry operation. In this way one recovers
a set of  quantum numbers associated with the original
symmetries of the Hamiltonian. Already at this SR  level genuine defects are induced in the intrinsic determinant
$| {\cal{D}} \rangle$ resulting from the VAP procedure. Therefore, its
structure  differs from the one obtained
within the standard HF approximation. \cite{non-unitary-paper-Carlos}
This kind of SR symmetry-projected
framework has already enjoyed considerable success in
quantum chemistry. \cite{PQT-reference-1,PQT-reference-2,PQT-reference-3}

An extension of the SR method \cite{Carlo-review}
has been previously used \cite{Rayner-2D-Hubbard-PRB-2012} to  describe  half-filled and doped  2D Hubbard lattices. With the help of chains
of VAP calculations, it provides a (truncated) basis consisting
of a few (orthonormalized) symmetry-projected configurations which can then be
used  to further diagonalize the considered  Hamiltonian. In this way, one can account
on an equal footing
for additional correlations in both ground and excited states keeping well
defined symmetry quantum numbers. The method also
provides a well-controlled ansatz to compute both spectral functions (SFs) and
density of states (DOS). \cite{Rayner-2D-Hubbard-PRB-2012} In quantum
chemistry, the first benchmark calculations
on the $C_{2}$ dimer have  shown that, with a modest basis set, this
methodology provides a high quality description of the low-lying
spectrum for the entire dissociation profile. In addition, the same methodology
has been applied to obtain the full low-lying spectrum of formaldehyde
as well as to a challenging model $C_{2 \nu}$ insertion pathway for
BeH$_{2}$. \cite{Carlos-Rayner-Gustavo-VAMPIR-molecules}

%
%
\begin{figure}
\includegraphics[width=0.50\textwidth]{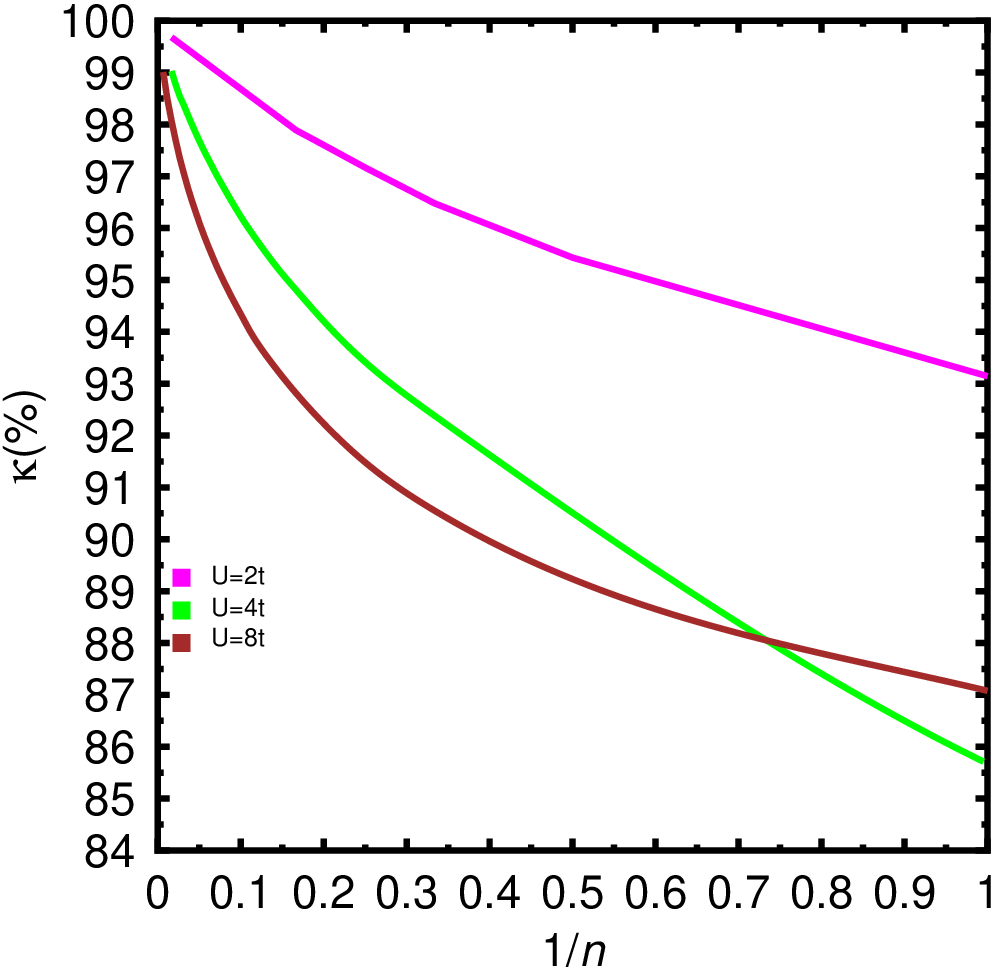}
\caption{(Color online) The ratio of correlation energies $\kappa$ obtained with the FED approximation
is plotted as a function of the inverse of the number of GHF transformations for a  lattice
with $N_{sites}$=30 sites and $N_{e}$=22 electrons. Results are shown for on-site repulsions of $U$=2$t$, 4$t$, and 8$t$.
For details, see the main text.
}
\label{en-vs-tra}
\end{figure}
%
%

However, being already more sophisticated than the SR framework, \cite{Carlos-Hubbard-1D,PQT-reference-1,PQT-reference-2,PQT-reference-3}
the extension \cite{Rayner-2D-Hubbard-PRB-2012,Carlos-Rayner-Gustavo-VAMPIR-molecules}  mentioned above
still essentially describes a given ground and/or excited state in terms
of a single symmetry-projected configuration. This certainly limits the amount
of correlations that can be accessed for those  states.
A more correlated description is encoded in a MR scheme.
\cite{Tomita-1,Tomita-2,Tomita-2011PRB,rayner-Hubbard-1D-FED2013}
Here, one resorts to a set of symmetry-broken  HF states
$| {\cal{D}}^{i} \rangle$ and superposes their
Goldstone manifolds  $\hat{R}| {\cal{D}}^{i} \rangle$. In this way, a given
state with a well defined set of quantum numbers is expanded
in terms of  $n$ nonorthogonal symmetry-projected
configurations \cite{Tomita-1,rayner-Hubbard-1D-FED2013}
that are optimized with the help of the Ritz variational
principle \cite{Blaizot-Ripka}  applied to the projected energy.

There
are several ways to perform the
self-consistent
optimization of the intrinsic HF
states  $| {\cal{D}}^{i} \rangle$ within a MR approach. One possible VAP
strategy is represented by the Resonating HF
\cite{Tomita-1,Tomita-2,Tomita-2011PRB,Fukutome-original-RSHF,Yamamoto-1,Yamamoto-2,Ikawa-1993} (ResHF) scheme
within which
all the  determinants $| {\cal{D}}^{i} \rangle$ are optimized at the same time. Another
VAP strategy is represented by the 
Few  Determinant \cite{rayner-Hubbard-1D-FED2013,Carlo-review}
(FED)   approach
where, the  HF
transformations $ {\cal{D}}^{i}$
 are optimized one-at-a-time.
In both the ResHF and FED schemes, the corresponding configuration
mixing coefficients are determined through resonon-like equations. \cite{Gutzwiller_method}
We note that there is no need for the FED expansion to be short, as its name
would imply, although this is a desirable feature. In the present study, we keep the acronym to
remain consistent with the literature. \cite{Carlo-review}
Hybrid MR approximations are also possible. For example, one could optimize $n-k$
states using the ResHF scheme and $k$ states using the FED one.

%
%
\begin{figure*}
\includegraphics[width=0.90\textwidth]{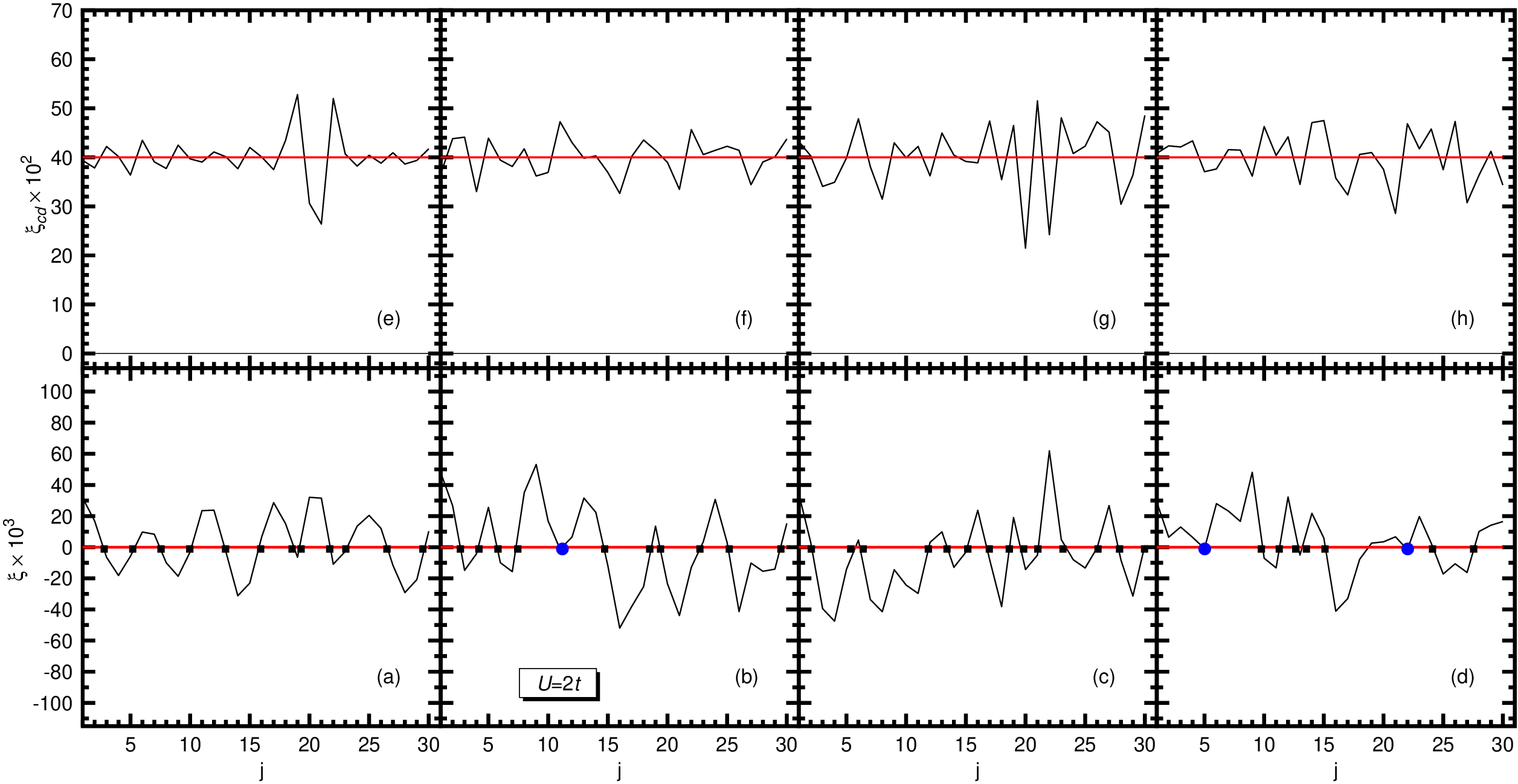}
\caption{(Color online) The quantities  ${\xi}^{i}(j)$
[panels (a) to (d)]
 and ${\xi}_{cd}^{i}(j)$ [(e) to (h)]
are plotted
for a 1D  lattice with $N_{sites}$=30 sites and $N_{e}$=18 electrons  at $U$=2$t$,  as a
  function of
lattice site $j$ for some typical symmetry-broken GHF determinants  resulting from the FED VAP optimization.
Results corresponding to  the standard RHF approximation, are plotted in red for comparison. For more details, see
the main text.
}
\label{solitons-N30-Ne18-U2}
\end{figure*}
%
%

The
 FED  methodology has already been used \cite{rayner-Hubbard-1D-FED2013}  to compute
 ground state energies, spin-spin correlation functions
 (SSCFs) in real space, magnetic structure factors (MSFs) as well as spin-charge separation
 tendencies in the SFs
 of half-filled 1D Hubbard lattices
 of different sizes
 in the weak, intermediate-to-strong,
 and strong interaction regimes.
 We have shown  \cite{Carlos-Rayner-Gustavo-FED-molecules}
 that short ResHF and FED
 expansions can provide an accurate
 description of chemical systems such as the nitrogen and water molecules
 along the entire dissociation profile,
 as well as  an accurate interconversion profile among the
 peroxo and bis($\mu$-oxo) forms of [Cu$_{2}$O$_{2}$]$^{2 +}$
 comparable to other state-of-the-art quantum chemical methods.
 Recent calculations, \cite{Laimis-paper}
 have also considered
 the complex binding pattern in the Mo$_{2}$ molecule.

  In addition to ground state properties, the
Excited Few Determinant \cite{Carlo-review} (EXCITED FED)
 scheme has also
 been used \cite{rayner-Hubbard-1D-FED2013} to treat excited states, with well defined quantum numbers, as
 expansions in terms of nonorthogonal symmetry-projected configurations. As
 a byproduct of  VAP calculations, the EXCITED FED provides a (truncated)
 basis consisting of a few Gram-Schmidt  orthonormalized states, each of them
 expanded in a given number of nonorthogonal
  symmetry-projected configurations, which may
 be used to perform a final diagonalization of the Hamiltonian 
 to account for a more correlated description of both ground
 and excited states.

Let us stress, that each of
the MR approaches already mentioned
has its
own advantages and drawbacks. A ResHF wave function
 is stationary with respect to arbitrary changes in the HF
transformations \cite{rs} ${\cal{D}}^{i}$ (i=1, $\dots$, $n$) while a FED one
displays stationarity only with respect to the last added transformation.
Therefore, the ResHF wave functions become easier to work with in the
evaluation  of those properties depending on derivatives of the wave function. However, in a
ResHF optimization, ${\cal{O}}(n^{2})$ Hamiltonian and norm kernels have to be
recomputed at every iteration while only ${\cal{O}}(n)$ kernels are required
in an efficient implementation of the FED method. \cite{rayner-Hubbard-1D-FED2013,Carlos-Rayner-Gustavo-FED-molecules}

Regardless of the FED and/or ResHF VAP strategy adopted, the MR
approximations  are not restricted by the dimensionality (i.e., they can be equally well applied 
to 1D and 2D systems)
and/or the topology of the considered lattices.  On the other hand, one of the most attractive
features of the MR  approximations is that they offer compact
wave functions, with well defined quantum numbers, whose quality
can be systematically improved
by increasing the number of symmetry-projected configurations included in the
corresponding
ansatz.
\cite{Tomita-1,Tomita-2,Tomita-2011PRB,rayner-Hubbard-1D-FED2013,Carlos-Rayner-Gustavo-FED-molecules}
Obviously, one is always limited in practical applications to a finite number $n$ of symmetry-projected
terms in  the FED and/or ResHF expansions. However, it should also be kept in mind that
both the ResHF and FED wave functions are nothing else than a discretized form of the exact coherent state representation
of a fermion state \cite{Perlemov}  and, therefore, become exact in the limit
$n \rightarrow \infty$.
All in all, we believe that symmetry-projected approximations,
already quite successful in nuclear
physics, \cite{rs,Carlo-review,rayner-GCM-paper,rayner-GCM-parity} lead to a rich conceptual landscape
and deserve further attention
in  quantum chemistry \cite{PQT-reference-1,PQT-reference-2,PQT-reference-3,Carlos-Rayner-Gustavo-VAMPIR-molecules,Carlos-Rayner-Gustavo-FED-molecules}
and condensed matter physics.
\cite{Carlos-Hubbard-1D,Rayner-2D-Hubbard-PRB-2012,non-unitary-paper-Carlos,rayner-Hubbard-1D-FED2013}
They also provide \cite{PaperwithShiwei} high quality trial states that can be used
within the  constrained-path Monte Carlo  \cite{CPMC-ref} scheme,
increasing the energy accuracy and decreasing the 
statistical
variance as more symmetries are broken and restored.

In this paper we apply, for the first time, the FED approach to 
doped systems. Therefore, our main goal is to test its 
performance using benchmark calculations. To this end, we have 
selected the 1D Hubbard model for which both exact and highly
accurate density matrix renormalization
group (DMRG) results can be obtained. For the sake of completeness
and comparison we will also discuss half-filling results.
In Sec. \ref{Theory}, we briefly describe the key
ingredients of our MR approach. For a more detailed account, the reader is referred
to our previous works. \cite{rayner-Hubbard-1D-FED2013,Carlos-Rayner-Gustavo-FED-molecules}
In Sec. \ref{results}, we discuss the results of our calculations. We have first
paid attention to
lattices with $N_{sites}$=30 sites and $N_{e}$=14, 18, 22, 26, 30 electrons
as illustrative examples. Calculations have been performed for the on-site repulsions
$U$=2$t$, 4$t$, and 8$t$ representing the weak, intermediate-to-strong
(i.e., noninteracting band width),
and strong
interaction regimes, respectively.
In Sec. \ref{GSandCoEnergies}, we compare our ground state
and correlation energies with the exact ones as well as
with those obtained using other theoretical methods. The
dependence of the correlation energies predicted for doped
lattices with the number of  transformations included in our MR
ansatz is also discussed in the same section. The basic units
of quantum fluctuations in the case of doped  lattices
are discussed
in Sec. \ref{UnitsQF}, where we consider the structure of the
symmetry-broken  determinants resulting from the FED VAP procedure.
Next, in Secs. \ref{MoDistribution}
and \ref{CorreFunctions}, we benchmark our results for momentum
distributions, SSCFs, and density-density (DDCFs)
correlation
functions with DMRG
ones obtained with the open-source
ALPS software. \cite{ALPS}
A typical outcome of our calculations for SFs and DOS
${\cal{N}}(\omega)$
is
presented in Sec. \ref{spectral}, where we consider
a lattice with $N_{sites}$=30 sites and $N_{e}$=26 electrons
at $U$=4$t$. Next, in Sec. \ref{larger-sites}, we illustrate
the performance of the FED method in the case of
larger lattice sizes.
Finally, Sec. \ref{conclusions}
is devoted to the concluding remarks and work perspectives.

%
%
\begin{figure*}
\includegraphics[width=0.90\textwidth]{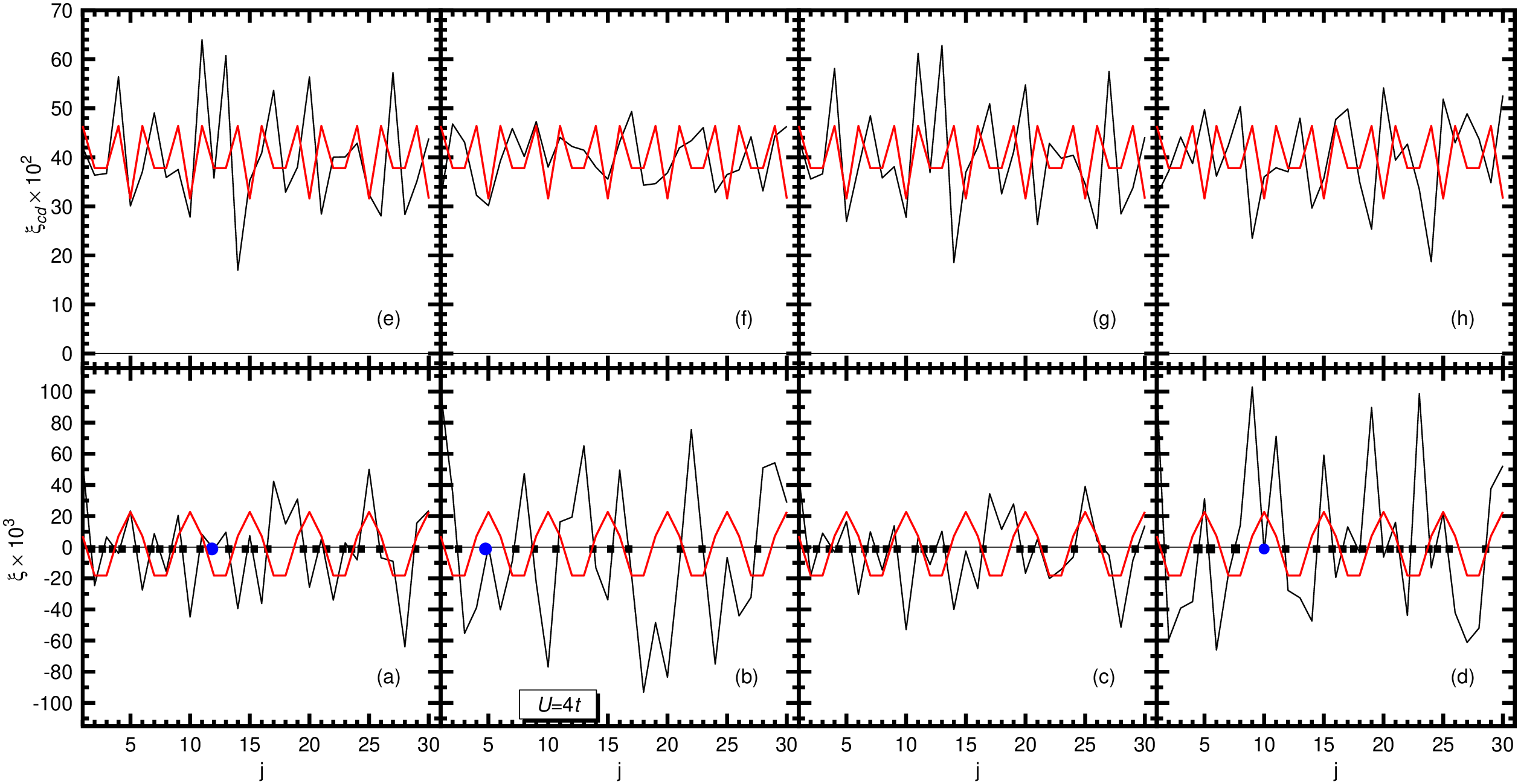}
\caption{(Color online)
Same as Fig. \ref{solitons-N30-Ne18-U2}
but for $U$=4$t$. Results corresponding to  the standard UHF approximation, are plotted in red for comparison.
}
\label{solitons-N30-Ne18-U4}
\end{figure*}
%
%

\section{Theoretical Framework}
\label{Theory}

We consider the  1D Hubbard Hamiltonian \cite{Hubbard-model_def1}

\begin{eqnarray} \label{HAM-hubbard1D}
\hat{H} =
-t \sum_{j,\sigma= \uparrow, \downarrow}
\Big \{
\hat{c}_{j+1 \sigma}^{\dagger} \hat{c}_{j \sigma}
+
\hat{c}_{j \sigma}^{\dagger} \hat{c}_{j+1 \sigma}
\Big \}
+
U \sum_{j} \hat{n}_{j \uparrow} \hat{n}_{j \downarrow}
\end{eqnarray}
where the first term represents the nearest-neighbor
hopping  (t $>$ 0) and the second is  the repulsive on-site  interaction (U $>$ 0).
The fermionic \cite{Blaizot-Ripka}  
spin-1/2
operators $\hat{c}_{j \sigma}^{\dagger}$ and $\hat{c}_{j \sigma}$
create and destroy an electron with spin-projection
$\sigma= \uparrow, \downarrow$ on a lattice site j=1, $\dots$, $N_{sites}$.
The operators
$\hat{n}_{j \sigma}$ = $\hat{c}_{j \sigma}^{\dagger} \hat{c}_{j \sigma}$  are the local number operators.
We assume periodic boundary conditions and a lattice  spacing $\Delta$=1.

The starting point \cite{rayner-Hubbard-1D-FED2013}
of our FED approach
is a set of
GHF determinants\cite{StuberPaldus,HFclassification}
$| {\cal{D}}^{i} \rangle$ ($i=1, \dots$, n), which deliberately
break
several symmetries of the Hamiltonian
 like rotational
(in spin space) and spatial ones.
To restore these broken
 symmetries, we explicitly use
the  spin
\begin{eqnarray} \label{PROJ-S}
\hat{P}_{\Sigma {\Sigma}^{'}}^{S} =
\frac{2S+1}{8 {\pi}^{2}} \int d \Omega {\cal{D}}_{\Sigma {\Sigma}^{'}}^{S *} (\Omega) R(\Omega)
\end{eqnarray}
and  space group

\begin{eqnarray} \label{proj-space-group}
\hat{P}_{m m^{'}}^{k} = \frac{h}{L} \sum_{g=1}^{L}
{\Gamma}_{m m^{'}}^{k *}(g) \hat{R}(g)
\end{eqnarray}
projection operators. \cite{rayner-Hubbard-1D-FED2013} In Eq.(\ref{PROJ-S}), $ R(\Omega)$ is
 the rotation operator in spin space, the label $\Omega$ stands for the set of Euler
angles, and ${\cal{D}}_{\Sigma {\Sigma}^{'}}^{S}(\Omega)$ are
Wigner matrices.  \cite{Edmonds}
In Eq.(\ref{proj-space-group}), $h$ and $L$ represent the dimension of the irreducible
representation and the number of space group operations for a given lattice.
On the other hand,  ${\Gamma}_{m m^{'}}^{k}(g)$
is an  irreducible representation \cite{Tomita-1,Lanczos-Fano}
while $\hat{R}(g)$ represents the corresponding point group symmetry operations
parametrized in terms of the label g. The linear momentum
${k}=\frac{2 \pi}{N_{sites}} {\xi}$ is given in terms of the quantum number
$\xi$
which
takes the values allowed inside the Brillouin zone (BZ). \cite{Ashcroft-Mermin-book}
The high symmetry momenta $k$=0,$\pi$ are also labelled by
the parity
of the corresponding irreducible representation. \cite{Tomita-1,Lanczos-Fano}
In what follows we will
not explictly write this label.
The total projection operator
can then be written in the following shorthand form

%
%
\begin{figure*}
\includegraphics[width=0.90\textwidth]{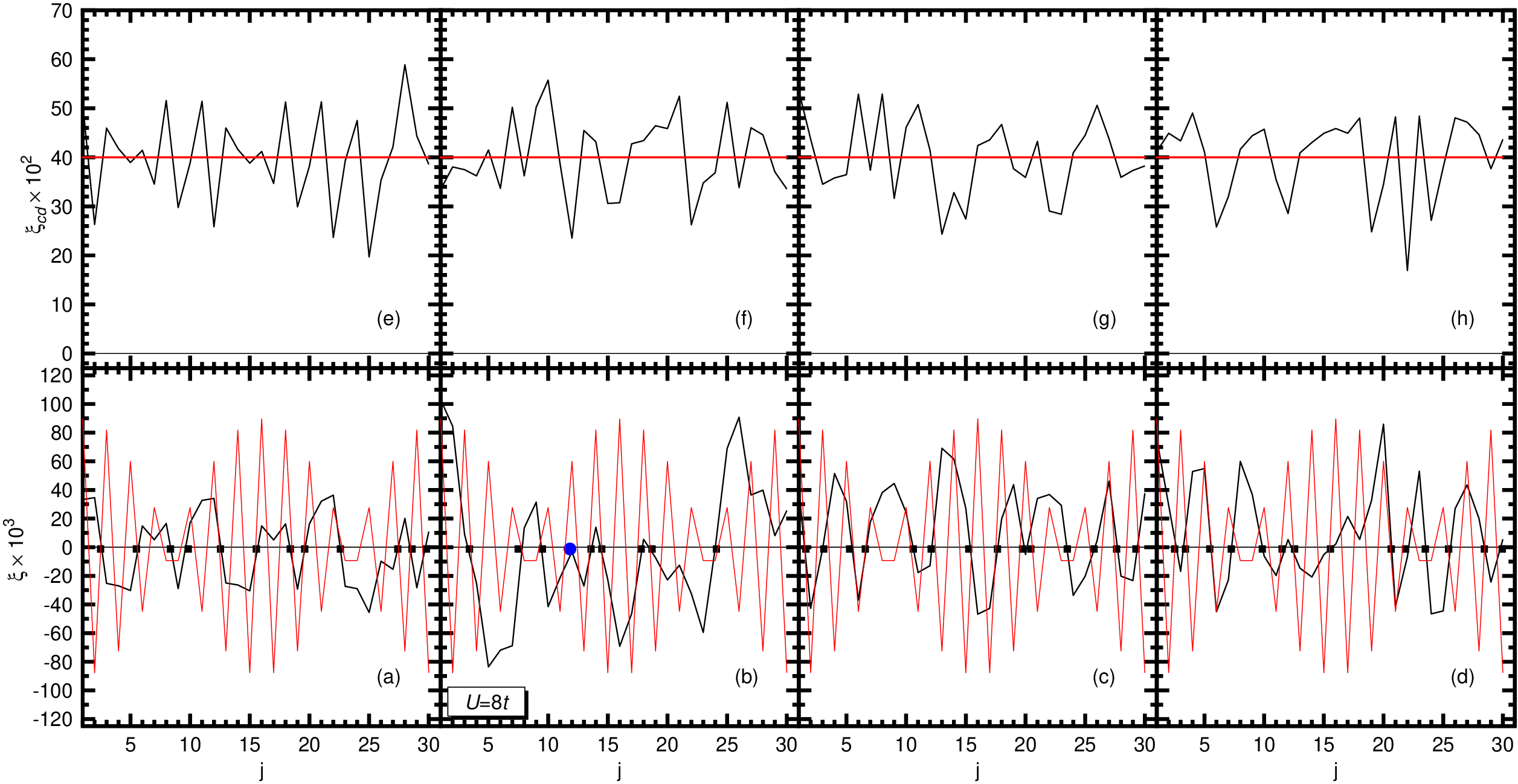}
\caption{(Color online)
Same as Fig. \ref{solitons-N30-Ne18-U2}
but for $U$=8$t$. Results corresponding to  the standard GHF approximation, are plotted in red for comparison.
}
\label{solitons-N30-Ne18-U8}
\end{figure*}
%
%

\begin{eqnarray} \label{projection-operator-P}
\hat{P}_{\Sigma  {\Sigma}^{'} }^{S }
\hat{P}_{m m^{'}}^{k}
= \hat{P}_{K K^{'}}^{\Theta}
\end{eqnarray}
where  $\Theta = (S,k)$
represents the set of
(spin  and linear momentum)
symmetry quantum numbers and $K=(\Sigma,m)$.
The key idea of the FED approach is  to
superpose the set of degenerate  Goldstone states \cite{rayner-Hubbard-1D-FED2013}
$| {\cal{D}}^{i} (\Omega,g)\rangle = \hat{R}(\Omega) \hat{R}(g) | {\cal{D}}^{i} \rangle$
through the following ansatz

\begin{eqnarray} \label{FED-state-general}
| \phi_{K}^{n \Theta} \rangle =
\sum_{K^{'}}
\sum_{i=1}^{n}
f_{K^{'} }^{i \Theta}
\hat{P}_{K K^{'}}^{\Theta}
| {\cal{D}}^{i} \rangle
\end{eqnarray}
which expands a given ground state $| \phi_{K}^{n \Theta} \rangle$, with well defined
symmetry quantum numbers $\Theta$, in terms of $n$ nonorthogonal
symmetry-projected configurations $\hat{P}_{K K^{'}}^{\Theta}| {\cal{D}}^{i} \rangle$.
The sum over $K^{'}$ in Eq.(\ref{FED-state-general}) is necessary in order to
remove an unphysical dependence of  $| \phi_{K}^{n \Theta} \rangle$
on the orientation of the GHF states $| {\cal{D}}^{i} \rangle$.
\cite{Schmid-Gruemmer-1984}

The FED wave function is  formally similar to the
one adopted within the ResHF method. \cite{Tomita-1,Tomita-2,Tomita-2011PRB,Fukutome-original-RSHF,Yamamoto-1,Yamamoto-2,Ikawa-1993}
It is determined by
applying the Ritz variational
principle \cite{Blaizot-Ripka} to the energy
(independent of K)

\begin{eqnarray} \label{ojo-ojo}
E^{n \Theta} =
\frac{
f^{n \Theta \dagger}
{\cal{H}}^{n \Theta}
f^{n \Theta}
}
{
f^{n \Theta \dagger}
{\cal{N}}^{n \Theta}
f^{n \Theta}
}
\end{eqnarray}
written in terms of Hamiltonian and norm kernels
\begin{eqnarray} \label{HNKernels-GHF-FED}
{\cal{H}}_{i K,j K^{'}}^{n \Theta} &=&
\langle {\cal{D}}^{i} | \hat{H} \hat{P}_{K K^{'}}^{\Theta}  | {\cal{D}}^{j} \rangle
\nonumber\\
{\cal{N}}_{i K,j K^{'}}^{n \Theta} &=&
\langle {\cal{D}}^{i} |  \hat{P}_{K K^{'}}^{\Theta}  | {\cal{D}}^{j} \rangle
\end{eqnarray}
which require the knowledge of the
symmetry-projected matrix elements
between all the GHF determinants
used in the expansion
Eq.(\ref{FED-state-general}). In the case of the mixing coefficients, we obtain a
resonon-like \cite{Gutzwiller_method} eigenvalue
equation

\begin{eqnarray} \label{HW-1}
\left({\cal{H}}^{n \Theta}
- E^{n \Theta}
{\cal{N}}^{n \Theta}
\right)
f^{\Theta} = 0
\end{eqnarray}
with the constraint
$f^{n \Theta \dagger} {\cal{N}}^{n \Theta} f^{n \Theta} = 1$ ensuring
the normalization of the solution. Within the
FED approach the energy Eq.(\ref{ojo-ojo}) is varied only
with respect to the last added GHF determinant
$| {\cal{D}}^{n} \rangle$ keeping all the other transformations
${\cal{D}}^{i}$ (i=1, $\dots$, $n-1$), obtained in previous chains of VAP calculations, fixed.
\cite{rayner-Hubbard-1D-FED2013} Note that at variance with the
ResHF approximation, \cite{Tomita-1,Tomita-2,Tomita-2011PRB,Fukutome-original-RSHF,Yamamoto-1,Yamamoto-2,Ikawa-1993}
where all the transformations $ {\cal{D}}^{i}$
are optimized at the same time, the FED VAP
strategy optimizes them one-at-a-time. This is
 particularly relevant for alleviating our numerical effort
 if one keeps in mind that we use the most general
 symmetry-broken GHF states and therefore a full 3D spin projection
 Eq.(\ref{PROJ-S})
 is required. Regardless of the
 adopted FED and/or ResHF strategy, the
 variation with respect to the transformations  ${\cal{D}}^{i}$
 can be efficiently parametrized with the help of the
 Thouless theorem.
\cite{Rayner-2D-Hubbard-PRB-2012,Carlos-Hubbard-1D,non-unitary-paper-Carlos,rayner-Hubbard-1D-FED2013,Carlos-Rayner-Gustavo-VAMPIR-molecules,Carlos-Rayner-Gustavo-FED-molecules}

 All the FED calculations discussed  in this paper have been carried out
with an in-house  parallel implementation
\cite{rayner-Hubbard-1D-FED2013}
of our VAP
procedure. We have used a limited-memory
quasi-Newton method \cite{quasi-Newton}
to handle the optimization.
Note that the FED expansion of a given
ground state
$| \phi_{K}^{n \Theta} \rangle$
 by $n$ nonorthogonal
symmetry-projected GHF configurations enlarges
the flexibility in our wave functions, with respect to a SR
description, to a total number
$n_{var}=2n(2N_{sites}-N_{e})\times N_{e}+ 4nS + 2(n-1)$
of variational parameters.

In Sec. \ref{spectral}, we will also discuss both
the SFs and DOS  ${\cal{N}}(\omega)$. The key point,  is to
superpose the  Goldstone hole
$| {\cal{D}}_{h}^{i} \left( \Omega, g\right) \rangle=
\hat{R}(\Omega) \hat{R}(g) {\hat{b}}_{h} \left({\cal{D}}^{i}\right) | {\cal{D}}^{i} \rangle$
and particle
$| {\cal{D}}_{p}^{i} \left( \Omega, g\right) \rangle=\hat{R}(\Omega) \hat{R}(g) {\hat{b}}_{p}^{\dagger} \left({\cal{D}}^{i}\right) | {\cal{D}}^{i} \rangle$
(i= 1, $\dots$, $n_{T}$)
manifolds  in the
wave functions of the
$N_{e}-1$ and $N_{e}+1$ electron systems, respectively. The amplitudes of these superpositions
are then determined through the corresponding  generalized eigenvalue equations similar
to Eq.(\ref{HW-1}). With these ingredients at hand, as well as the  FED solution
$| \phi_{K}^{n \Theta} \rangle$, the SFs and DOS can be computed according to
Eqs.(25) and (26) in our previous work.\cite{rayner-Hubbard-1D-FED2013}

%
%
\begin{figure}
\includegraphics[width=0.420\textwidth]{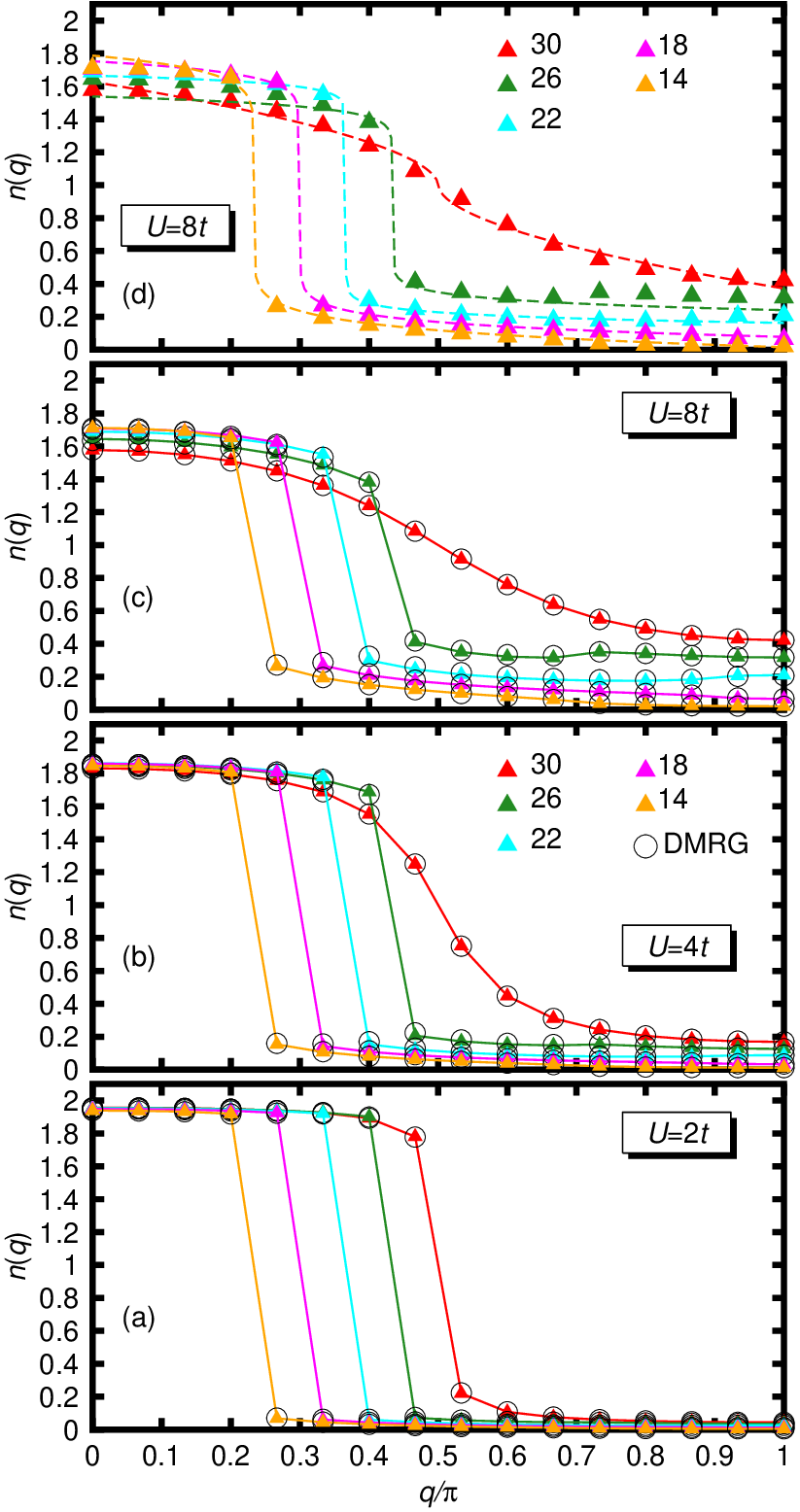}
\caption{(Color online) The ground state momentum distributions Eq.(\ref{nkdist})
for  $N_{sites}$=30 lattices  with 14 (orange triangles), 18 (magenta triangles), 22 (cyan triangles), 26
(green triangles), and
30 (red triangles) electrons
are shown for $U$=2$t$ (a), 4$t$ (b), and 8$t$ (c). DMRG results (open black circles) are also
included in  panels (a) to (c)  for comparison.
The FED values (filled triangles) are compared
in panel (d)
with a power-law [Eq.(\ref{power-law-kdist})]
fitting (dashed lines) of the momentum distributions.
}
\label{momentum-distribution}
\end{figure}
%
%

Finally, for the convenience of the reader we summarize the acronyms used in the present study 
for different types
of Slater determinants:

\begin{itemize}

\item RHF is used for those symmetry-adapted states preserving
all the symmetries of the  Hamiltonian Eq.(\ref{HAM-hubbard1D}).

\item UHF is used for those  states preserving the $\hat{S}_{z}$-symmetry while possibly breaking all others.

\item GHF states are those that break all the symmetries of the considered
Hamiltonian.

\end{itemize}

\section{Discussion of results}
\label{results}

In this section, we discuss the results of our FED calculations.
First, we pay attention to lattices with $N_{sites}$=30 sites and $N_{e}$=14, 18, 22, 26, 30
electrons. Results are presented
for  on-site repulsions $U$=2$t$, 4$t$, and 8$t$, respectively.
In Sec. \ref{GSandCoEnergies}, we compare the   ground state and
correlation energies with the exact ones, as well as with results obtained
using other theoretical approaches. We also discuss, for the case of doped
lattices, the dependence of the  correlation energies on the number
$n$ of nonorthogonal symmetry-projected GHF configurations. Next, in
Sec. \ref{UnitsQF}, we  consider
the structure of the intrinsic GHF determinants resulting from
our VAP procedure in the case of doped lattices. The momentum
distributions  are presented
in Sec. \ref{MoDistribution} while the Fourier transforms
of the SSCFs and  DDCFs in real space are shown in
Sec. \ref{CorreFunctions}. They are
compared with those obtained within the DMRG framework
retaining 1024 states in the renormalization procedure.
On the other hand, in Sec. \ref{spectral}, we discuss spin-charge
separation tendencies in the SFs and DOS ${\cal{N}}(\omega)$
of a lattice with
$N_{sites}$=30 sites and $N_{e}$=26 electrons. Finally, in Sec. \ref{larger-sites}, we illustrate
the performance of the FED method in the case of
larger lattices.

\subsection{Ground state and correlation energies}
\label{GSandCoEnergies}

In Table \ref{Table1}, we compare the
exact \cite{LIEB,BETHE} and the
predicted FED ground state energies for
half-filled and doped lattices with $N_{sites}$=30 sites
and $N_{e}$=14, 18, 22, 26, 30 electrons.
The corresponding $\Theta$ = (0,0) ground
states have A$_{1}$ symmetry, i.e., they are symmetric under the
reflection $x \rightarrow -x$. For $U$=2$t$ and 4$t$, the FED energies
shown in the table have  been obtained by including $n$=60 nonorthogonal  symmetry-projected  GHF
configurations in the ansatz Eq.(\ref{FED-state-general}).
On the other hand, $n$=150 GHF transformations have been used for
$U$=8$t$. Besides the RHF 
 energies, we have  included in Table  \ref{Table1}
the lowest possible HF solution for completeness.
At $U$=2$t$, the  lowest-energy HF solution coincides with the RHF one
in the case of doped systems while an  UHF  solution is obtained
at half-filling. At $U$=4$t$ the
HF state corresponds to an UHF wave function  while for
$U$=8$t$ we have found a  GHF  solution with predominant ferromagnetic character.
In the same table, we also show  our previous results \cite{rayner-Hubbard-1D-FED2013} (FED$^{*}$), based on $n$=25 GHF transformations, and the ResHF ones \cite{Tomita-1}
obtained with $n$=30 UHF transformations. We have computed the ratio

\begin{eqnarray} \label{formulaCE}
\kappa_{FED} = \frac{E_{RHF}-E_{GHF-FED}}{E_{RHF}-E_{Exact}} \times 100
\end{eqnarray}
in order to check how well the FED correlation energies  reproduce the exact ones. For the other approximations, such a ratio is obtained
from a similar expression.

%
%
\begin{figure}
\includegraphics[width=0.40\textwidth]{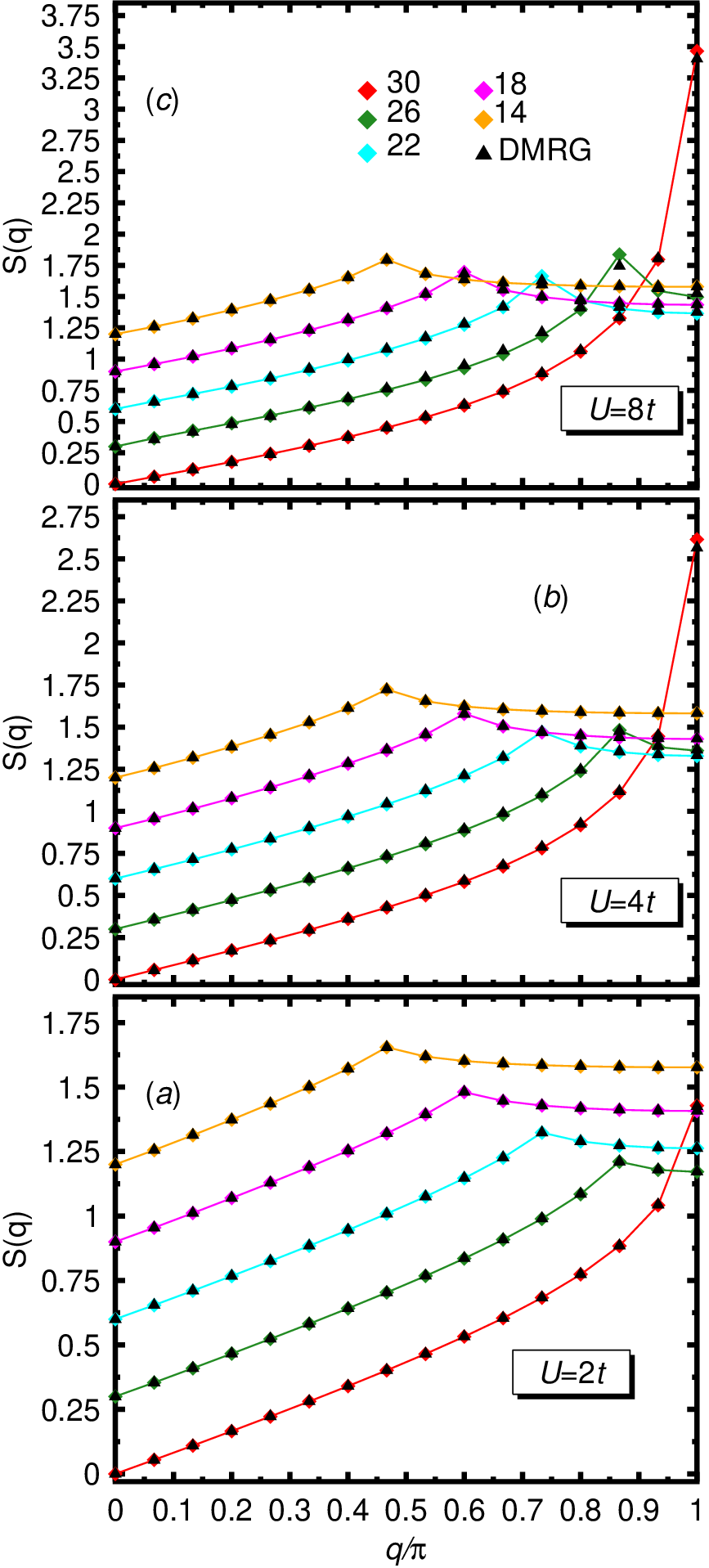}
\caption{(Color online)
Fourier transforms of the  ground state spin-spin correlation functions
in real space for  $N_{sites}$=30 lattices  with 14 (orange diamonds), 18 (magenta diamonds), 22
(cyan diamonds), 26
(green diamonds), and
30 (red triangles) electrons
are shown for $U$=2$t$ (a), 4$t$ (b), and 8$t$ (c). DMRG results (black triangles) are also
included   for comparison. Starting with 26 electrons, all the curves have
been successively shifted by 0.3 to accomodate them in a single plot.
}
\label{spinspin-momentum-space}
\end{figure}
%
%

The first noticeable feature from Table \ref{Table1} is that, at most, the standard
HF solutions account for 93.65 $\%$ of the exact correlation energy. Regardless of the filling and/or the
interaction strength, the MR FED expansion 
clearly recovers a very large portion of correlation energy (i.e., $\kappa_{FED}$ $\ge$   99 $\%$)
in all cases studied. For the half-filled case as well as for the lattices
with  26 and 22 electrons, the ground state and correlation energies
improve the ResHF  and FED$^{*}$ ones
obtained in  previous studies. \cite{Tomita-1,rayner-Hubbard-1D-FED2013}
On the other hand, the DMRG
energies (not  shown in the table)
are  exact to all the quoted figures.
We have further used the
DMRG results in Secs. \ref{MoDistribution}
and \ref{CorreFunctions}
 to
benchmark our calculations for
momentum distributions and
correlation functions.

From these  results and the ones obtained in our 
previous work, \cite{rayner-Hubbard-1D-FED2013}
 we conclude that
the  FED scheme provides a reasonable  starting point  to obtain correlated ground state
wave functions, with well defined symmetry quantum numbers, in both half-filled and
doped 1D Hubbard lattices. In addition, the method offers a systematic way to
improve, through chains of VAP calculations, the quality of such wave
functions by increasing the number $n$ of nonorthogonal symmetry-projected GHF
configurations included in the FED ansatz. This is illustrated in Fig. \ref{en-vs-tra} where
we have plotted, as a function of the inverse $1/n$ of the number  of
transformations, the ratio $\kappa_{FED}$ for
a doped lattice with $N_{sites}$=30 sites and $N_{e}$=22 electrons.
One sees that
$\kappa_{FED}$ increases smoothly and approaches the exact result as
the number of symmetry-projected configurations
is increased. For example, a
single symmetry-projected configuration provides
 $\kappa_{FED}$=93.14 $\%$, 85.67 $\%$, 87.08 $\%$  while increasing the number of GHF transformations
up to $n$=10 we obtain $\kappa_{FED}$=98.37 $\%$, 96.23 $\%$  and 94.35 $\%$
for
$U$=2$t$, 4$t$ and 8$t$, respectively.

Some comments are in order here. First, since the nature of the
quantum correlations varies for
different doping fractions
$x=N_{e}/N_{sites}$
and
on-site repulsions, one
can expect that the number of GHF transformations  required
to obtain a given $\kappa_{FED}$ ratio depends on both of them.
As already mentioned the FED wave functions become exact in the limit
$n \rightarrow \infty$. In practice we are always
limited to a finite number of nonorthogonal symmetry-projected
configurations in the FED expansion and it is difficult to assert beforehand how 
many of them are required. Therefore, their number $n$ should be
tailored, through chains of VAP calculations, so as to
reach
a reasonable accuracy not only in the ground state energy
 but also  in other  physical quantities like, for
example, the spin-spin correlators. In the present study we have used
a fixed number $n$=60 for both the weak and intermediate-to-strong interaction
regimes while a larger number $n$=150
is required to obtain the energies reported in Table \ref{Table1} at $U$=8$t$.
As we will see later on in Sec. \ref{CorreFunctions}, a larger
number of transformations is also required, especially
close to
half-filling, to
improve the quality
of the predicted  correlation functions. This can be qualitatively understood
from the crossover
in the SSCFs \cite{CrossOverSS} that explains how the antiferromagnetic
spin correlation at half-filling grows near  half-filling.
One may then expect strong
quantum fluctuations
near half-filling, whose basic units
(see, Sec. \ref{UnitsQF})
 can only be captured
with  larger FED expansions. The performance of the FED method for larger lattices
will be discussed  in Sec. \ref{larger-sites}.

\subsection{Structure of the intrinsic determinants and
basic units of quantum fluctuations in doped lattices}
\label{UnitsQF}

In our previous study \cite{rayner-Hubbard-1D-FED2013}
of the half-filled 1D Hubbard model we have considered two orders parameters, i.e.,  the
spin density (SD)

\begin{eqnarray} \label{SDW1}
\xi(j) = (-)^{j-1}
\langle {\cal{D}}  | \hat{\bf{S}}(j) | {\cal{D}} \rangle
\cdot
\langle {\cal{D}} | \hat{\bf{S}}(1) | {\cal{D}} \rangle
\end{eqnarray}
and the charge density (CD)

\begin{eqnarray} \label{ChargeDensity}
{\xi}_{cd}(j)= 1-\sum_{\sigma} \langle {\cal{D}} | \hat{n}_{j \sigma} | {\cal{D}} \rangle
\end{eqnarray}
associated with an arbitrary symmetry-broken determinant $| {\cal{D}} \rangle$, with
j=1, $\dots$, $N_{sites}$ being the lattice index.
The comparison
of the SD and CD computed with the standard UHF  solution
and the ones obtained using the GHF determinants
$| {\cal{D}}^{i} \rangle$  resulting from the FED VAP
procedure, reveals   that  ${\xi}^{i}(j)$ displays
neutral [i.e., ${\xi}_{cd}^{i}(j)$=0]
solitons \cite{Horovitz}
whose translational and breathing motions
can be regarded
\cite{Tomita-1,rayner-Hubbard-1D-FED2013,Yamamoto-1,Ikawa-1993}
as the basic units of
 quantum fluctuations in the FED wave functions Eq.(\ref{FED-state-general}).

 The question
 naturally arises, as to what are the basic units of the quantum fluctuations
 captured within the FED VAP optimization in the case of doped 1D  lattices.
 Among all the GHF  transformations $ {\cal{D}}^{i}$ used
 to describe the  lattice with $N_{sites}$=30  sites and $N_{e}$=18 electrons
 (see, Sec. \ref{GSandCoEnergies}), we have
 selected
 some typical
 examples to plot in Figs. \ref{solitons-N30-Ne18-U2}, \ref{solitons-N30-Ne18-U4}
 and \ref{solitons-N30-Ne18-U8}
 the corresponding SD ${\xi}^{i}(j)$
 [panels (a) to (d)]
 and  CD ${\xi}_{cd}^{i}(j)$ [panels (e) to (h)]
 as functions of lattice site. Results obtained with the lowest-energy
 standard HF solutions
 are also included
 in the plots (red) for comparison.

 In the case of the RHF solution (Fig.\ref{solitons-N30-Ne18-U2})
 at $U$=2$t$, the corresponding SD vanishes while the CD
 takes the constant value 0.4. Due to the
 symmetry-broken nature of the UHF  solution (Fig.\ref{solitons-N30-Ne18-U4}), the corresponding
 SD and CD exhibit  oscillating patterns around the expected
 values (i.e., 0 and 0.4, respectively) at $U$=4$t$. In the
 case of the intrinsic GHF solution (Fig.\ref{solitons-N30-Ne18-U8}) at $U$=8$t$
 the SD displays a very fast oscillating pattern, which is a direct
 consequence of its predominant ferromagnetic character
 [note the presence of the factor $(-)^{j-1}$ in the definition
 Eq.(\ref{SDW1})]. In fact, the energy of this intrinsic  GHF solution (i.e, -18.1922 t)
 is only slightly lower  than the one (i.e., -18.0974 t) corresponding to a fully
 ferromagnetic UHF solution with all the spins aligned along the z-direction.
 On the other hand, the CD takes the constant value 0.4.

Regardless of the considered interaction regime, the  GHF determinants
$| {\cal{D}}^{i}\rangle$ associated with the FED solution Eq.(\ref{FED-state-general})
exhibit pairs
of solitons
(black squares) where the SD ${\xi}^{i}(j)$ changes its sign. The 
space group projection operator provides a translational motion 
for such soliton pairs. When different  determinants show 
soliton pairs with different widths, this can be interpreted 
as a breathing mechanism. Isolated points (blue circles)
where   ${\xi}^{i}(j)$ becomes zero are also apparent  from
 Figs. \ref{solitons-N30-Ne18-U2}, \ref{solitons-N30-Ne18-U4}
 and \ref{solitons-N30-Ne18-U8}. In the case of
doped lattices these new defects represent polarons.  \cite{Tomita-1}
In addition, the CD  ${\xi}_{cd}^{i}(j)$ displays local variations around the
constant value 0.4 for all the considered on-site repulsions. Other GHF determinants
$| {\cal{D}}^{i}\rangle$ (not shown in the figures) display the same
qualitative features. Similar results also hold for other U
values and lattices.

Therefore, in the case of doped 1D lattices, the
 ansatz Eq.(\ref{FED-state-general}) superposes manifolds
$| {\cal{D}}^{i} \left( \Omega, g\right) \rangle=\hat{R}(\Omega) \hat{R}(g) | {\cal{D}}^{i} \rangle$ containing
both solitons and polarons. One is then left with an intuitive
physical picture
\cite{Tomita-1,rayner-Hubbard-1D-FED2013}
in which the basic units  of quantum fluctuations
in 1D lattices can be mainly associated with the translational and
breathing motions of neutral and charged solitons. However, in the
case of doped 1D systems, a part of such fluctuations can also be
described by polarons. Within
 both the FED \cite{rayner-Hubbard-1D-FED2013}
 and the ResHF \cite{Tomita-1,Yamamoto-1,Ikawa-1993} schemes, the
 interference between the defects belonging to different symmetry-broken
 determinants $| {\cal{D}}^{i} \rangle$ is accounted for through
  Eq.(\ref{HW-1}).

%
%
\begin{figure}
\includegraphics[width=0.4500\textwidth]{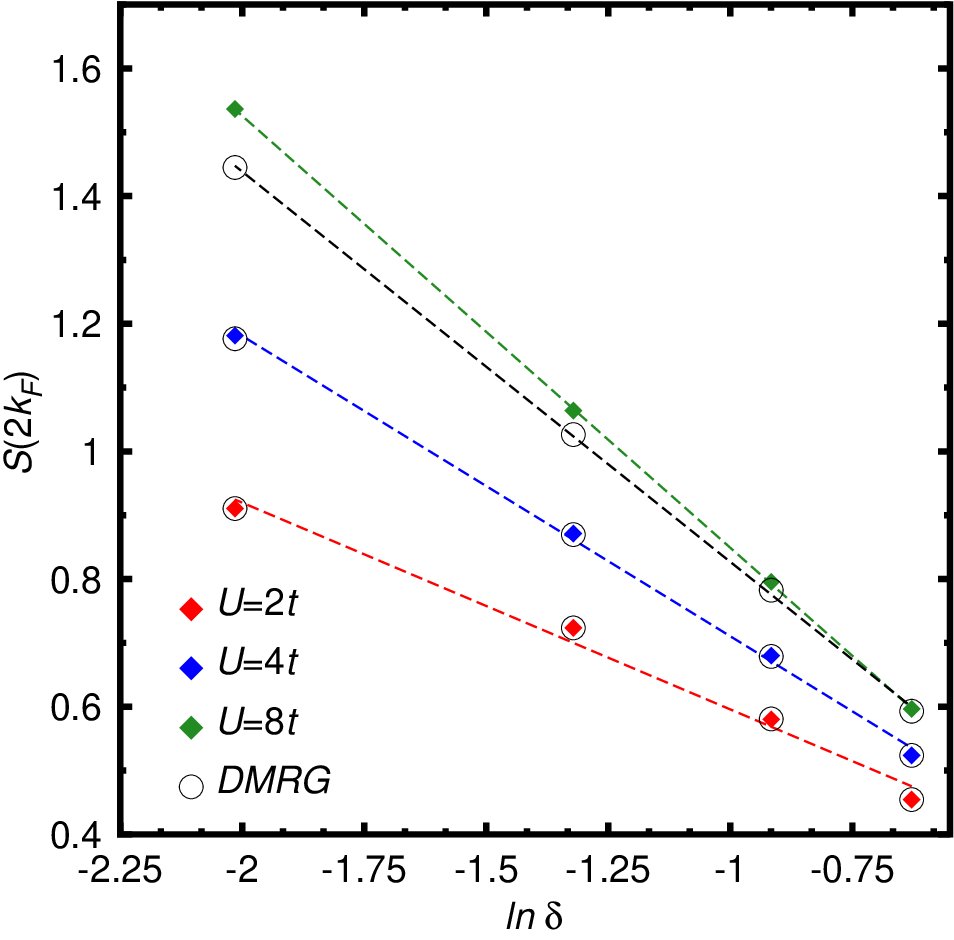}
\caption{(Color online) Maxima of Fourier transforms of the FED ground state spin-spin correlation functions
in real space
[Eq.(\ref{ss-CF})] plotted as functions of ln $\delta$, with
$\delta$ being the corresponding doping parameter.
Results
are shown for on-site repulsions $U$=2$t$ (red diamonds), 4$t$ (blue diamonds), and 8$t$ (green diamonds).
DMRG values are plotted with open circles. A straight line has been fitted to guide the eye.
For more details, see the main text.
}
\label{spinspin-maxima}
\end{figure}
%
%

\subsection{Momentum distribution}
\label{MoDistribution}

For a given set of quantum numbers $\Theta$, the momentum distribution
$n^{n \Theta}(q)$ can be computed as

\begin{eqnarray} \label{nkdist}
n^{n \Theta}(q) = \sum_{\sigma}
\frac{
\langle \phi_{K}^{n \Theta} | {\hat{n}}_{q \sigma} | \phi_{K}^{n \Theta} \rangle
}
{
\langle \phi_{K}^{n \Theta}  | \phi_{K}^{n \Theta} \rangle
}
\end{eqnarray}
where $\hat{n}_{q \sigma}$ is the $\sigma$-occupation operator at wave vector q.
Note that, due to the particular form of the  operator
$\hat{n}_{q \sigma}$, the momentum distribution Eq.(\ref{nkdist})
does not depend explicitly on $K$.

The ground state momentum distributions
for $N_{sites}$=30 lattices with
14 (orange triangles), 18 (magenta triangles), 22 (cyan triangles), 26
(green triangles), and
30 (red triangles) electrons are plotted in panels
(a), (b) and (c) of Fig. \ref{momentum-distribution} for on-site
repulsions $U$=2$t$, 4$t$, and 8$t$, respectively. Regardless of the interaction
strength, the FED and DMRG (open black circles) momentum distributions
agree well. In all cases, we have obtained  a  jump at $q=k_{F}$, with $k_{F}$ being
the Fermi momentum, that becomes less pronounced, especially at half-filling, for larger
U values. Such a jump
is
also found in calculations based on the
exact
solution at $U$=$\infty$ \cite{Ogata-Shiba}
as well as  in previous studies. \cite{Sorella-1990,Qin-S2kf}
As can seen from panel (c), the momentum distribution
presents a slight nonmonotonic behavior close to half-filling
(i.e., $N_{e}$=26) due to a small feature near $q=2k_{F}$.

Previous works
\cite{Ogata-Shiba,Sorella-1990,Qin-S2kf,Solyom,SchulzPRL1990}
have shown that, contrary to an ordinary Fermi liquid, the
momentum distribution of the 1D Hubbard model exhibits a power-law
behavior
around $q=k_{F}$ given by

\begin{eqnarray} \label{power-law-kdist}
n(q) = n(k_{F}) + C |q-k_{F}|^{\tau} \text{sgn} \left(q-k_{F} \right)
\end{eqnarray}
at half-filling or U $\to$ $\infty$. We have used the momentum distributions obtained with the FED
approach  to fit the functional
dependence Eq.(\ref{power-law-kdist}). The results are shown in panel
(d) of Fig. \ref{momentum-distribution} for $U$=8$t$. Despite the fact that
the functional form Eq.(\ref{power-law-kdist}) has been obtained
using the exact $U$=$\infty$ solution, we observe that it nicely
reproduces the trends in the FED (and also DMRG) results.

\subsection{Correlation functions}
\label{CorreFunctions}

Let us now turn our attention to the predicted FED SSCFs and DDCFs.
We compare them with the corresponding DMRG values
in the case of
$N_{sites}$=30 lattices with
14 (orange triangles), 18 (magenta triangles), 22 (cyan triangles), 26
(green triangles), and
30 (red triangles) electrons. This comparison
will allow us to reveal to which extent the FED scheme
can capture the main short, medium and long range features
in these correlators, especially
in the case of doped lattices. We have resorted to the momentum space representation
(i.e., the Fourier transforms) of the  SSCFs and  DDCFs in real space.
For a discussion of the predicted FED SSCFs in
the case of half-filled lattices, the reader is also
referred to our previous work. \cite{rayner-Hubbard-1D-FED2013}

%
%
\begin{figure}
\includegraphics[width=0.4000\textwidth]{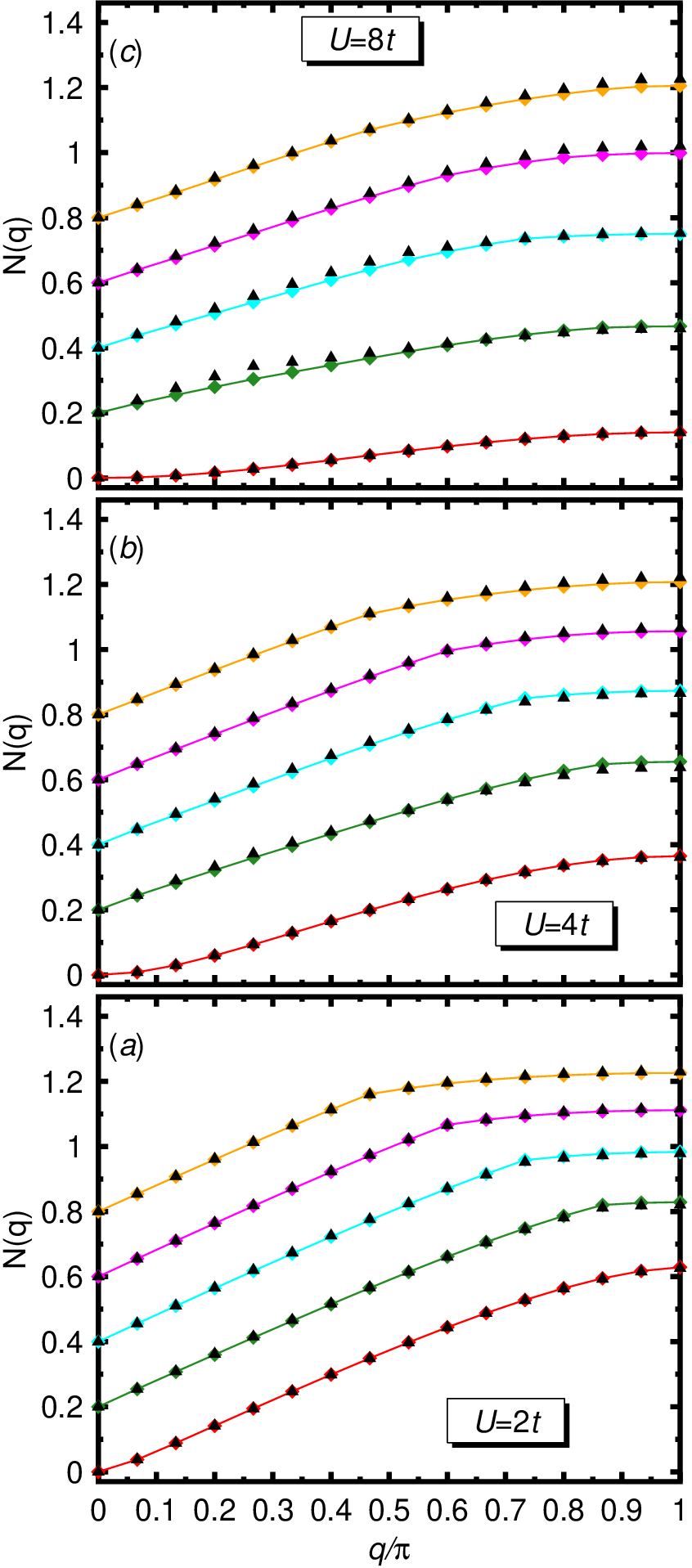}
\caption{(Color online) Same as Fig. \ref{spinspin-momentum-space}
but for the Fourier transforms of the  ground state density-density correlation functions
in real space.
Starting with 26 electrons, all the curves have
been successively shifted by 0.2 to accomodate them in a single plot.
}
\label{denden-momentum-space}
\end{figure}
%
%

The SSCFs in real space are given by

\begin{eqnarray} \label{ss-CF}
S_{m}^{n \Theta}(j) =
\frac{
\langle \phi_{K}^{n \Theta} | \hat{\bf{S}}(j) \cdot \hat{\bf{S}}(1) | \phi_{K}^{n \Theta} \rangle
}
{
\langle \phi_{K}^{n \Theta}  | \phi_{K}^{n \Theta} \rangle
}
\end{eqnarray}
where the subindex $m$   accounts for the
dependence with respect to the particular row  of the space group
irreducible representation used in the projection.
Let us stress that the wave functions
$| \phi_{K}^{n \Theta} \rangle $
Eq.(\ref{FED-state-general}) are pure spin states
where orbital relaxation is allowed. Both conditions have already been shown to be
important ingredients to improve the description of the long-range
behavior of the SSCFs. \cite{Tomita-1,rayner-Hubbard-1D-FED2013,Yamamoto-1,Ikawa-1993}
The Fourier transforms (FT-SSCFs) $S_{m}^{n \Theta}(q)$ of the SSCFs Eq.(\ref{ss-CF})
are depicted in panels (a), (b) and (c) of Fig. \ref{spinspin-momentum-space}
for the ground states of the lattices considered in the present study at
 $U$=2$t$, 4$t$ and 8$t$, respectively.

The first feature apparent from Fig. \ref{spinspin-momentum-space}, is the prominent
antiferromagnetic peak at the wave vector q=$\pi$ in the case of the half-filled system.
The peaks of the FT-SSCFs always occur at $q=2k_{F}$.
Such peaks have also been found \cite{Ogata-Shiba} with the exact
$U$=$\infty$ solution  of the 1D Hubbard model as well as in previous
calculations. \cite{Hirsch-S2kf,Qin-S2kf,Imada-S2kf}
They are shifted towards smaller linear momenta
as we move away from half-filling. In the same plot, we have
also included the results of our DMRG calculations (black triangles) for comparison.
It is satisfying to observe that the predicted FED values closely follow the trend obtained
within the DMRG approach. The largest differences between the FED and DMRG FT-SSCFs arise 
in the values of the corresponding peaks near half-filling for large U. For the lattice
with 26 electrons the FED peak, obtained with $n$=150 GHF transformations,  at $U$=8$t$
overestimates the DMRG one
by 5 $\%$. On the other hand, using a smaller number  $n$=60 of transformations, we
 obtain a poorer (10 $\%$ overestimation)
  description of the SSCFs and FT-SSCFs
in the strong interaction regime.

%
%
\begin{figure}
\includegraphics[width=0.4500\textwidth]{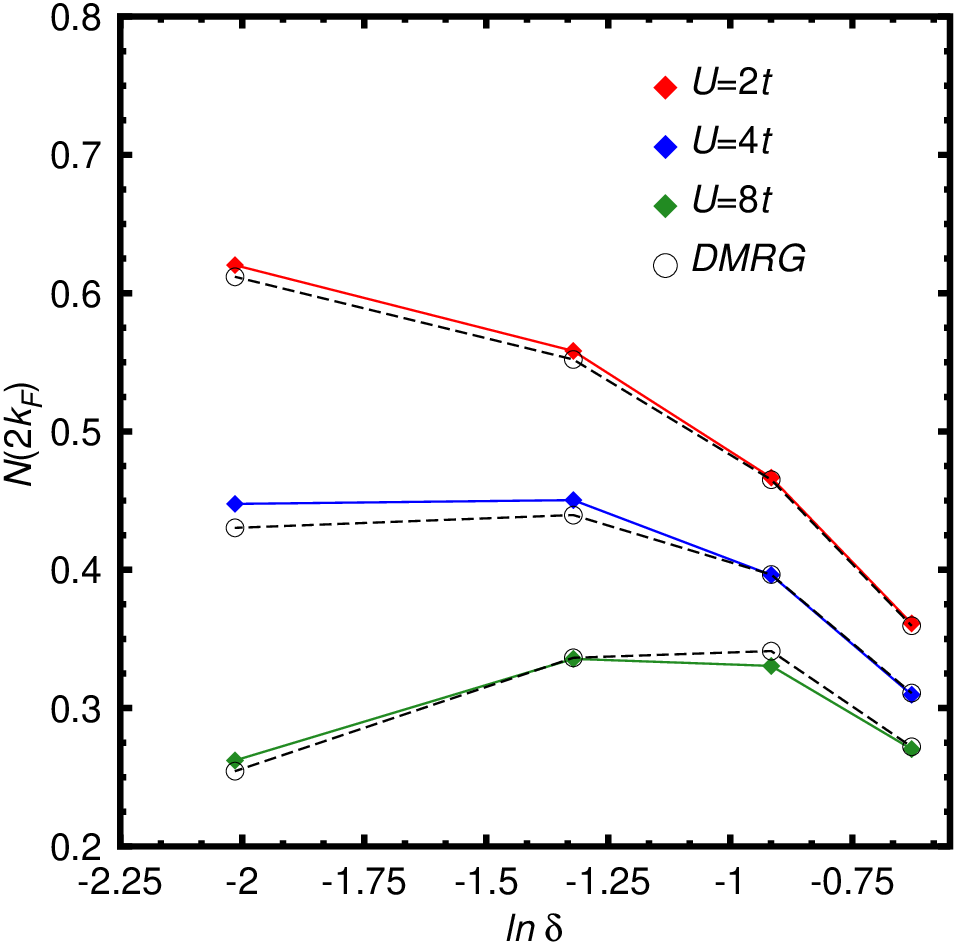}
\caption{(Color online)
Inflection points at wave vector $q=2k_{F}$  of the FED  Fourier-transformed density-density correlation functions
in real space
[Eq.(\ref{denden-CF})] are plotted as functions of ln  $\delta$ for $U$=2$t$ (red diamonds), 4$t$ (blue diamonds), and 8$t$ (green diamonds).
DMRG values are plotted with open circles.
For more details, see the main text.
}
\label{denden-maxima}
\end{figure}
%
%

A previous study \cite{CrossOverSS} has shown
the universal character
of the crossover in the SSCFs as we approach half-filling. As a result of this crossover, the
peaks observed in the FT-SSCFs at  $q=2k_{F}$ display
a linear logarithmic dependence with the doping parameter
$\delta=1 - x$. The results of our FED calculations
are compared in Fig. \ref{spinspin-maxima} with the DMRG ones.
For each U value, we have fitted a straight line to
guide the eye.  For $U$=2$t$ and 4$t$, the FED and DMRG
values agree well (we have therefore included only
the fitting of the  former in the plot) and
exhibit an almost
linear behavior as a function of  ln $\delta$. The same linear
trend is observed at $U$=8$t$ though in this case the
discrepancy between the FED and DMRG values, arising from a
poorer description of the
FT-SSCFs in the former (see, Fig. \ref{spinspin-momentum-space}), is larger.

%
%
\begin{figure*}
\includegraphics[width=0.80\textwidth]{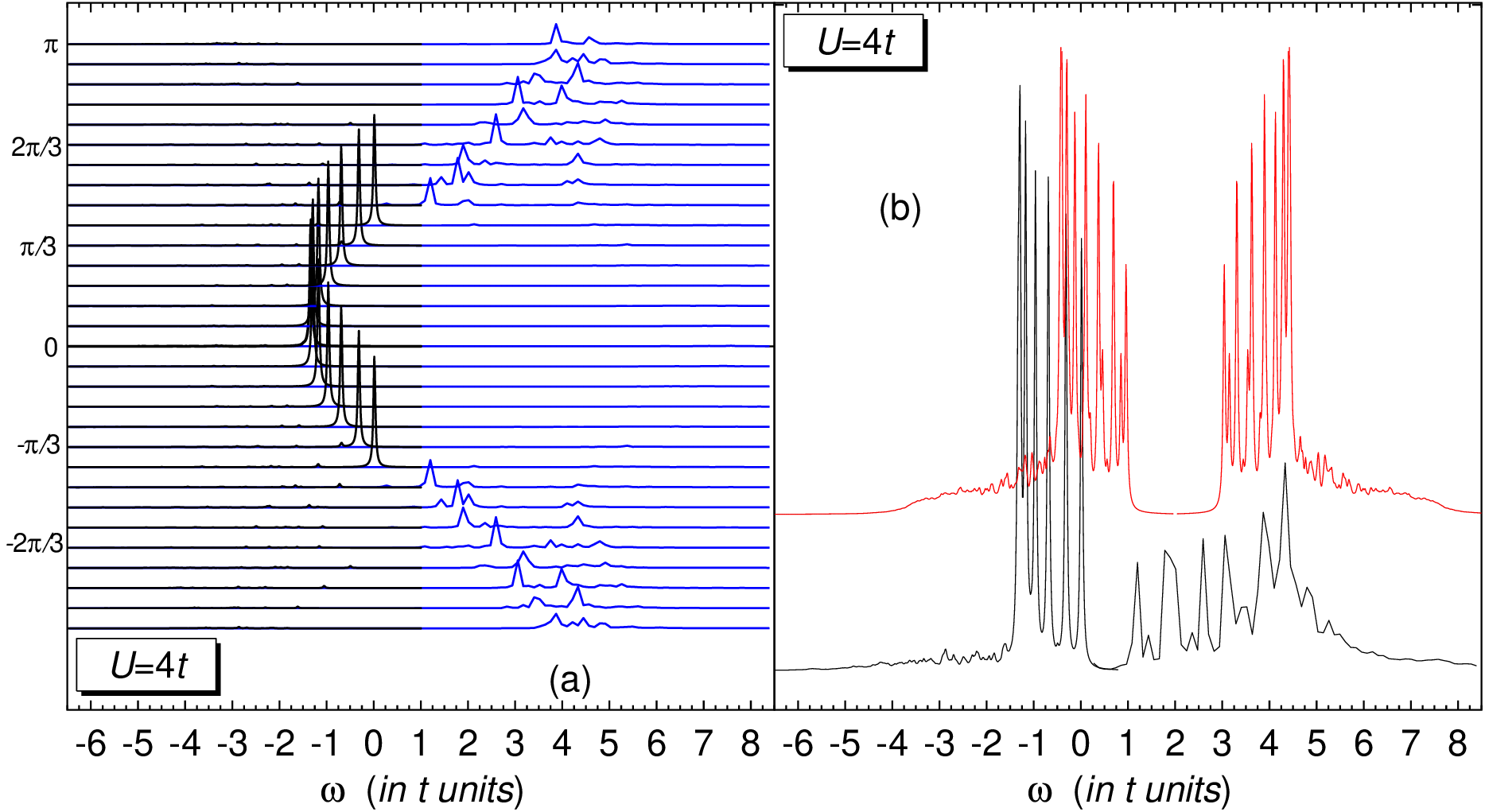}
\caption{(Color online) The hole (black) and particle (blue) SFs for
a lattice with
$N_{sites}$=30 sites and $N_{e}$=26
electrons
 are plotted
in panel (a) as functions of the  excitation energy $\omega$ (in t units).
The  DOS (black) ${\cal{N}}(\omega)$ is compared in panel (b) with the one (red)
corresponding to half-filling. The latter has been shifted
(i.e., ${\cal{N}}(\omega) + 40$)
for the sake of clarity.
Results are shown for the on-site repulsion $U$=4$t$.
A Lorentzian folding of width
$\Gamma$=0.05 t has been used.
For more details, see the main text.
}
\label{spectral-DOS}
\end{figure*}
%
%

The DDCFs in real space
can be computed as

\begin{eqnarray} \label{denden-CF}
N_{m}^{n \Theta}(j) =
\frac{
\langle \phi_{ K}^{n \Theta} | \delta \hat{n}(j) \delta \hat{n}(1) | \phi_{K}^{n \Theta} \rangle
}
{
\langle \phi_{K}^{n \Theta}  | \phi_{K}^{n \Theta} \rangle
}
\end{eqnarray}
where $\delta \hat{n}(j)= \hat{n}(j) - \langle \hat{n}(j) \rangle$
and $\hat{n}(j)= \sum_{\sigma} \hat{n}_{j \sigma}$.
Their  FT-DDCFs $N_{m}^{n \Theta}(q)$ are  shown in
panels (a), (b), and (c) of Fig. \ref{denden-momentum-space}
for
 $U$=2$t$, 4$t$, and 8$t$, respectively. DMRG values  (black triangles)
 are also plotted
  for comparison.
Similar to the momentum distributions and FT-SSCFs already discussed above, the
FED  FT-DDCFs closely follow the trends observed in the DMRG ones, with the largest
differences arising for $q \le \pi/2$ in the case of the lattice with 26 electrons at
$U$=8$t$. From  panels (a), (b), and (c) of Fig. \ref{denden-momentum-space}, we also
observe the appearence of inflection points
around  $q=2k_{F}$
which become less pronounced as U increases. They are plotted in Fig. \ref{denden-maxima} as functions
of  ln $\delta$. We observe a shift down of the curves as U increases reflecting
that the charge fluctuations decrease for larger U values. In addition, we note
that the curves bend down more for larger on-site repulsions.

\subsection{Spectral functions and density of states}
\label{spectral}

A typical outcome of our calculations is shown in  panel (a) of
Fig. \ref{spectral-DOS}, where we have plotted the hole (black) and particle (blue) SFs, as functions of
the excitation energy $\omega$, for a
lattice with $N_{sites}$=30 sites and $N_{e}$=26 electrons  at $U$=4$t$.
Calculations
have been performed along the lines described in our previous work. \cite{rayner-Hubbard-1D-FED2013}
The FED ground state of the system with $N_{e}$=26 electrons has been approximated by $n$=60
GHF transformations while for the systems with
$N_{e} \pm 1$ electrons we have superposed
$n_{T}$=25 hole and particle manifolds (see, Sec. \ref{Theory}). Smaller  values $n_{T}$=5 and $n_{T}$=15
have also been investigated. However, increasing the number of hole and particle manifolds to
$n_{T}$=25 leads to a shift of the main peaks and a redistribution of the strength of some of the
peaks found for  $n_{T}$=5 and $n_{T}$=15 as a result of the small number of configurations used
in the calculations. In all cases, a Lorentzian folding of width
$\Gamma$=0.05 t has been used.

We observe prominent hole peaks belonging to the spinon band. Such a band resembles the one
found in the half-filled case \cite{rayner-Hubbard-1D-FED2013} though the spectral weight of some of the hole peaks found in the latter is  redistributed
 to the
particle sector due to the presence of doping. Another prominent feature of the
 SFs shown in panel (a) is the very extended
distribution of the spectral weight for linear momenta $|k| > k_{F}$.
The splitting of the strength in the corresponding particle SFs
reveals that the present finite size results at the intermediate-to-strong
interaction
regime already display spin-charge separation tendencies beyond
a simple quasiparticle distribution as well
as shadow features.
 This agrees well with results obtained
using other theoretical approximations. \cite{AraGo,Senechal-scs,PENcPRL1996}

In panel (b) of Fig. \ref{spectral-DOS} we have plotted the DOS
${\cal{N}}(\omega)$
corresponding to
the  lattice with $N_{sites}$=30 sites and $N_{e}$=26 electrons (black). For the sake
of comparison, we have also included in the same panel the DOS in the half-filled
case (red). The last one, has also been computed using $n$=60 and $n_{T}$=25. It exhibits
the characteristic Hubbard gap \cite{moukouri2001,huscroft2001,aryanpour2003,AraGo}
and particle-hole symmetry. \cite{text-Hubbard-1D} As can be seen from the figure, this
particle-hole symmetry is lost in the doped case. As a result of states intruding the
original gap, a smaller pseudogap is developed at $U$=4$t$. However, our calculations
indicate that such a pseudogap progressively disappears for increasing
doping fractions $x$. Similar results also hold for both $U$=2$t$ and $U$=8$t$ though the effect
is less pronounced in the latter due to the larger value of the gap at half-filling.

\begin{table}
\label{Table2}
\caption{The ground state energies predicted within the FED scheme are
compared with the exact ones. Results are presented for
$N_{sites}$=34 lattices with $N_{e}$=14, 18, 22, 26, 30, 34 electrons as well as for the half-filled
lattice with  $N_{sites}$=50 sites. In each case, the number $n$ of transformations used
in the FED ansatz [Eq.(\ref{FED-state-general})] is indicated. The ratio of correlation energies $\kappa$, is
computed according to Eq.(\ref{formulaCE})
}
\begin{tabular}{ccccccccccccc}
\hline
\\
$N_{sites}$  & & $N_{e}$   & & U             & &    FED      &  & $n$	&  &  Exact   & & $\kappa$ ($\%$) \\
\\
\hline
34           & & 14        & &  4$t$           & & -23.0991    &  & 60	&  & -23.1137 & & 99.46	   \\
\\
34           & & 14        & &  4$t$           & & -23.1048    &  & 100	&  & -23.1137 & & 99.67	   \\
\\
34           & & 18        & &  4$t$           & & -26.4440    &  & 60	&  & -26.4842 & & 98.99	   \\
\\
34           & & 18        & &  4$t$           & & -26.4587    &  & 100	&  & -26.4842 & & 99.35	   \\
\\
34           & & 22        & &  4$t$           & & -27.8402    &  & 60	&  & -27.9207 & & 98.48	    \\
\\
34           & & 22        & &  4$t$           & & -27.8673    &  & 100	&  & -27.9207 & & 98.99	    \\
\\
34           & & 26        & &  4$t$           & & -27.1248    &  & 60	&  & -27.2553 & & 98.05	    \\
\\
34           & & 26        & &  4$t$           & & -27.1634    &  & 100	&  & -27.2553 & & 98.63	    \\
\\
34           & & 30        & &  4$t$           & & -24.2555    &  & 60	&  & -24.3967 & & 98.28	    \\
\\
34           & & 30        & &  4$t$           & & -24.2966    &  & 100	&  & -24.3967 & & 98.78	     \\
\\
34           & & 34        & &  4$t$           & & -19.4646    &  & 60	&  & -19.5258 & & 99.40	    \\
\\
34           & & 34        & &  4$t$           & & -19.4876    &  & 100	&  & -19.5258 & & 99.62	    \\
\\
34           & & 34        & &  8$t$           & & -11.0562    &  & 60	&  & -11.1473 & & 99.74	   \\
\\
34           & & 34        & &  8$t$           & & -11.0883    &  & 100   &  & -11.1473 & & 99.83          \\
\\
50           & & 50        & &  2$t$           & & -42.1748    &  & 60    &  & -42.2443 & & 98.46                \\
\\
50           & & 50        & &  2$t$           & & -42.1957    &  & 100   &  & -42.2443 & & 98.62              \\
\\
50           & & 50        & &  4$t$           & & -28.1924    &  & 60	&  & -28.6993 & & 96.62	   \\
\\
50           & & 50        & &  4$t$           & & -28.2999    &  & 100	&  & -28.6993 & & 97.33 	 \\
\\
50           & & 50        & &  8$t$           & & -15.7739    &  & 60    &  & -16.3842 & &  98.84         \\
\\
50           & & 50        & &  8$t$           & & -15.9770    &  & 100   &  & -16.3842 & &  99.22         \\
\hline
\end{tabular}
\end{table}

\subsection{Larger lattices}
\label{larger-sites}

Now, we turn our attention to larger lattices. Let us stress that
our aim in this section is not to be exhaustive but to illustrate
the FED results in such lattices. To this end the predicted ground state energies are
compared with the exact ones in Table II.
Results are presented for
$N_{sites}$=34 lattices with $N_{e}$=14, 18, 22, 26, 30, 34 electrons as well as for the half-filled
lattice with  $N_{sites}$=50 sites. For the considered on-site repulsions we have performed
two different sets of FED calculations based on $n$=60 and $n$=100 GHF transformations.
The ratio of correlation energies $\kappa$, has been
computed according to Eq.(\ref{formulaCE}).

As can be seen from the table, the FED approach, based on $n$=60 symmetry-projected
configurations, already provides ${\kappa}_{FED}$ $>$ 98 $\%$. These values
significantly
improve the ones obtained with the standard HF approximation. For example, for the
half-filled lattice with $N_{sites}$=50, the UHF approximation  accounts for
${\kappa}_{UHF}$=12.02 $\%$, 65.02 $\%$, 92.26 $\%$
while ${\kappa}_{FED}$= 98.46 $\%$, 96.62 $\%$, 98.84 $\%$
at $U$=2$t$, 4$t$, and 8$t$, respectively.
Note that, for the same on-site repulsions, the variational Monte Carlo
method provides $\kappa$ values of
around 87 $\%$, 92 $\%$, and 96 $\%$, respectively. \cite{Yokoyamma-Shiba}
For the same half-filled lattice, the ground state energies obtained
with $n$=60 also improve the values reported in our previous work \cite{rayner-Hubbard-1D-FED2013}
using a smaller number (i.e., $n$=25) of nonorthogonal
symmetry-projected GHF configurations in the FED expansion Eq.(\ref{FED-state-general})
as well as the ones provided by the ResHF approximation \cite{Tomita-1} based on $n$=30 UHF
transformations. Moreover, further increasing up to $n$=100 leads  to
${\kappa}_{FED}$ $\ge$ 99.35 $\%$ in the case of lattices with $N_{sites}$=34 sites.
In the case $N_{sites}=N_{e}$=50, we have obtained
${\kappa}_{FED}$= 98.62 $\%$, 97.33 $\%$, 99.22 $\%$ for $U$=2$t$, 4$t$, and 8$t$, respectively.
The previous results show that the FED scheme also provides a resonable starting point
to obtain correlated wave functions in lattices larger than the ones considered
in Sec. \ref{GSandCoEnergies}. In particular, one sees that for increasing
lattice sizes the quality of the FED wave functions can also be systematically
improved, for different doping fractions, by
increasing the number of symmetry-projected configurations included
in the MR ansatz Eq.(\ref{FED-state-general}). We are unable at the moment
to anticipate the number of symmetry-projected configurations
necessary to achieve a given quality in the FED wave functions for
arbitrary lattice sizes and doping fractions. Nevertheless, we stress once more that
the exact answer can always be approached in a systematic constructive way.

\begin{figure}
\includegraphics[width=0.420\textwidth]{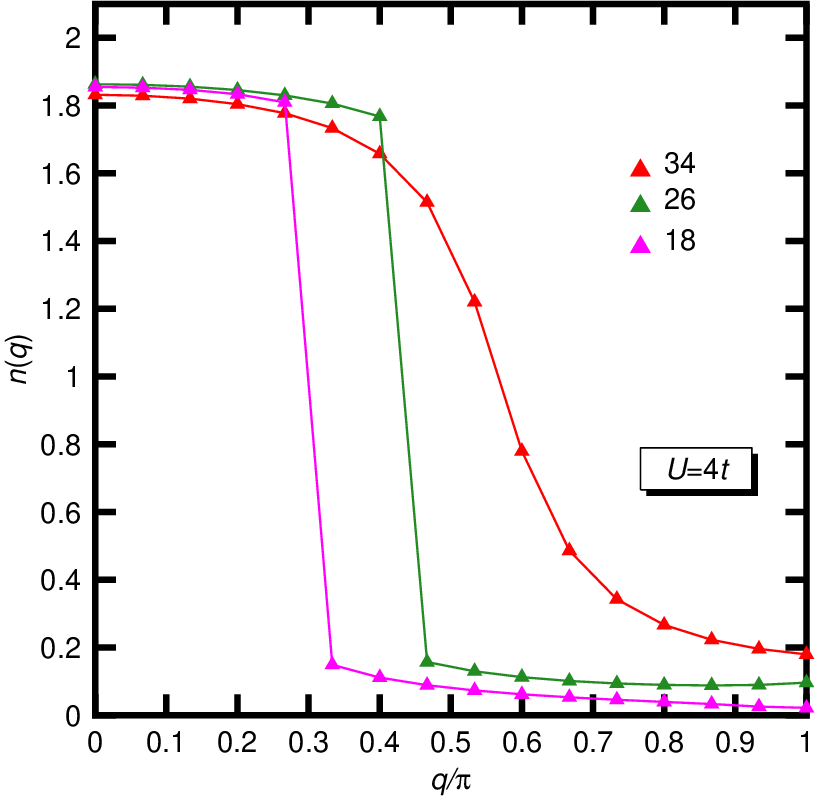}
\caption{(Color online) The ground state momentum distributions Eq.(\ref{nkdist})
for  $N_{sites}$=34 lattices  with  18 (magenta triangles), 26
(green triangles), and
34 (red triangles) electrons
are shown for $U$=4$t$.
}
\label{momentum-distribution-N34}
\end{figure}

Finally, as a typical example of the results obtained for the lattices considered in this section, we
have plotted in Fig. \ref{momentum-distribution-N34} the momentum distributions corresponding to $N_{sites}$=34 sites
with  18 (magenta triangles), 26
(green triangles), and
34 (red triangles) electrons
at $U$=4$t$. We have resorted to $n$=100 nonorthogonal symmetry-projected states in the
calculations. As can be seen from the figure, the momentum distributions still
display the main feature already discussed in Sec. \ref{MoDistribution}
for the case of half-filled and doped lattices with $N_{sites}$=30 sites, i.e., a jump
at $q=k_{F}$ that becomes less pronounced at half-filling.

%
%
%
%

\section{conclusions}
\label{conclusions}

In the present study we have applied, for the first time, the FED approach
to doped Hubbard systems. Our main goal has been to test its performance 
using benchmark calculations  in 1D Hubbard lattices.  Half-filled systems have also
been discussed. We have compared the results
of our calculations for ground state and correlation energies
with those obtained using other theoretical approximations. From
the results of our  previous study \cite{rayner-Hubbard-1D-FED2013}
and those obtained in the present work based on a larger
number of nonorthogonal symmetry-projected GHF configurations in the
MR expansion, we conclude that the FED scheme provides a reasonable
starting point to obtain (compact) correlated wave functions
in both
half-filled and doped 1D Hubbard  lattices. We have shown that the quality of such
wave functions can be systematically improved, through chains of
VAP calculations, in a constructive manner
by increasing the number of transformations in the corresponding FED ansatz.

The analysis of the  structure of the (intrinsic) symmetry-broken
Slater determinants resulting from our VAP procedure reveals
that they differ from that provided by the standard HF approximation. In
particular, in the case
of doped lattices they contain defects (i.e., solitons and polarons). The
translational and breathing motions of such  solitons can be regarded
as the  basic units of quantum fluctuations for the considered lattices.
In addition, in the case of doped 1D systems,  a part of the quantum fluctuations
can also be described by polarons. On the other hand, though the FED results are not
as accurate as the DMRG ones for the considered 1D lattices, our benchmark
calculations for momentum distributions and correlations functions show
that the former captures the main physics trends found in the latter.

We have also shown that the FED scheme can be used to access dynamical
properties of doped 1D Hubbard lattices such as SFs and the DOS. To this end, in
addition to the corresponding FED ground state based on $n$ GHF transformations, we
have considered
ans\"atze
for the $N_{e}+1$ and $N_{e}-1$ electron systems
that superpose $n_{T}$ particle and hole manifolds, respectively. For the
case of a doped lattice with $N_{sites}$=30 sites and $N_{e}$=26 electrons our
scheme provides hole and particle SFs that qualitatively
agree with results obtained using other theoretical frameworks. They point
point to a distribution of the spectral strength beyond the one expected
for a simple quasiparticle distribution and display spin-charge
separation tendencies in all the considered interaction regimes.

We believe that the finite size FED calculations already
show that VAP MR  expansions, based on
nonorthogonal symmetry-projected Slater determinants, represent a useful theoretical
tool to study low-dimensional correlated electronic systems with different
doping contents that complement other existing approaches and could even be
combined with them. Within this context,  we have recently used \cite{PaperwithShiwei}
SR symmetry-projected wave functions as trial states within the constrained-path
Monte Carlo framework. It has been shown that the use of such SR
symmetry-projected states increases the energy accuracy while decreasing
the statistical variance in calculations for large lattices. Given the fact that
short FED-like expansions encode a more correlated description of the considered
systems, they might be seen as plausible candidates for further improving the
previous results.

The MR expansions used in the present study still offer a rich conceptual
landscape for further development. In particular, small vibrations
around symmetry-projected mean fields (i.e., symmetry-projected
Tamm Dancoff and Random Phase approximations) can be consistently
formulated. \cite{Scmid-RPA,Nishiyama-RPA} Such approximations can then be used to access
a large number of excited states as required in studies
of the optical conductivity in lattice models. \cite{optical-Dagotoo,optical-Scalapino,Jeckelmann}
Such calculations are
in progress and will be reported elsewhere.

Let us stress that symmetry-projected approximations are not restricted
by the dimensionality of the considered lattices. In this respect, our
studies have paved the way for applying the MR methodology to
the systematic description of both ground and excited states of
2D square, honeycomb, Kagome, and triangular lattices,
as well as more involved multi-orbital Hubbard models
relevant to iron-based superconductors. \cite{Dagotto-Rev-Mod-Physics-2013}
In the realm of quantum chemistry, we plan to further enlarge
our current developments
\cite{Carlos-Rayner-Gustavo-VAMPIR-molecules,Carlos-Rayner-Gustavo-FED-molecules}
for the molecular Hamiltonian.

$
$

\begin{acknowledgments}

This work was supported
by the Department of Energy, Office of Basic Energy Sciences, Grant No. DE-FG02-
09ER16053. G.E.S. is a Welch Foundation Chair (C-0036).
Some of the calculations in this work
have been performed at the Titan computational facility, Oak
Ridge National Laboratory, National Center for Computational
Sciences, under project CHM048. The authors also acknowledge
a computational grant received from
the National Energy Research Scientific
Computing Center (NERSC) under the project Projected Quasiparticle Theory.
One of us (R.R-G.) would like to thank
Prof. K. W. Schmid, Institut f\"ur Theoretische Physik
der Universit\"at T\"ubingen, for valuable  discussions.

\end{acknowledgments}


\begin{thebibliography}{00}



\bibitem{Dagotto-review} E. Dagotto, Rev. Mod. Phys. {\bf{66}},  763 (1994).

\bibitem{Dagotto-Rev-Mod-Physics-2013} E. Dagotto, Rev. Mod. Phys. {\bf{85}}, 849 (2013).

\bibitem{HTCSC-1} J. G. Bednorz and K. A. M\"uller, Z. Phys. B {\bf{64}}, 189 (1986).

\bibitem{Hubbard-model_def1} J. Hubbard, Proc. R. Soc. (London)  A {\bf{276}}, 238 (1963).

\bibitem{Vollhardt} D. Vollhardt, Rev. Mod. Phys. {\bf{56}}, 99 (1984).

\bibitem{Anderson-1} P. W. Anderson, Science {\bf{235}}, 1196 (1987).


\bibitem{Science-Dagotto} E. Dagotto, Science {\bf{309}}, 257 (2005).

\bibitem{sup-Fe} Y. Kamihara, T. Watanabe, M. Hirano
and H. Hosono, J. Am. Chem. Soc. {\bf{130}}, 3296 (2008).

\bibitem{Stewart-Review} G. R. Stewart, Rev. Mod. Phys. {\bf{83}}, 1539 (2011).


\bibitem{Su-Fe-Dagotto} P. Dai, J. Hu and E. Dagotto, Nature {\bf{8}}, 709 (2012).




\bibitem{optical-1} R. J\"ordens, N. Strohmaier, K. G\"unter, H. Moritz and T. Esslinger, Nature
{\bf{455}}, 204 (2008); U. Schneider, L. Hackerm\"uller, S. Will, Th. Best, I. Bloch,
T. A. Costi, R. W. Helmes, D. Rasch and A. Rosch, Science {\bf{322}}, 1520 (2008);
I. Bloch, J. Dalibard and W. Zwerger, Rev. Mod. Phys. {\bf{80}}, 885 (2008).

\bibitem{CastroNeto-review} A. H. Castro Neto, F. Guinea, N. M. R. Peres, K., S. Novosolev and
A. K. Geim, Rev. Mod. Phys. {\bf{81}}, 109 (2009).

\bibitem{text-Hubbard-1D} F. H. L. Essler, H. Frahm, F. G\"ohmann, A. Kl\"umper
and V. E. Korepin, {\em The One-Dimensional Hubbard Model} (University Press, Cambridge, 2005).



\bibitem{Mikeska} H. -J. Mikeska and K. Kolezhuk, Lect. Notes Phys. {\bf{645}}, 1 (2004).




\bibitem{Lanczos-Fano} G. Fano, F. Ortolani and
 A. Parola, Phys. Rev. B {\bf{46}}, 1048 (1992).


\bibitem{Nightingale} {\it{Quantum Monte Carlo Methods in Physics and Chemistry}} edited
by M. P. Nightingale and C. J. Umrigar, NATO Advanced Studies Institute, Series C: Mathematical
and Physical Sciences (Kluwer,
Dordrecht, 1999), Vol. 525.


\bibitem{Raedt-MC} H. De Raedt and W. von der Linden, {\it{The Monte Carlo Method in Condensed Matter Physics}}, edited
by K. Binder (Springer-Verlag, Heidelberg, 1992).

\bibitem{Neuscamman-2012} E. Neuscamman, C. J. Umrigar and G. K.-L. Chan, Phys. Rev. B {\bf{85}}, 045103 (2012).


\bibitem{Bishop-1} R. F. Bishop, P. H. Y. Li, D. J. J. Farnell, J. Richter and C. E. Campbell, Phys. Rev. B {\bf{85}}, 205122 (2012).

\bibitem{Bishop-2} P. H. Y. Li, R. F. Bishop, D. J. J. Farnell, and C. E. Campbell, Phys. Rev. B {\bf{86}}, 144404  (2012).

\bibitem{var-red-den-mat} J. R. Hammond and D. A. Mazziotti, Phys. Rev. A {\bf{73}}, 062505 (2006).


\bibitem{DMRG-White} S. R. White, Phys. Rev. Lett. {\bf{69}}, 2863 (1992).

\bibitem{Dukelsky-Pittel-RPP} J. Dukelsky and S. Pittel, Rep. Prog. Phys. {\bf{67}}, 513 (2004).

\bibitem{Scholl-RMP} U. Schollw\"ock, Rev. Mod. Phys. {\bf{77}},  259 (2005).


\bibitem{Scholl-AP} U. Schollw\"ock, Ann. Phys. {\bf{326}}, 96 (2010).

\bibitem{GChan} G. K.-L. Chan and S. Sharma, Ann. Rev. Phys. Chem. {\bf{62}},  465 (2011).

\bibitem{TNPS-1} L. Tagliacozzo, G. Evenbly and G. Vidal, Phys. Rev. B {\bf{80}}, 235127 (2009).

\bibitem{TNPS-2} C. V. Kraus, N. Schuch, F. Verstraete and J. I. Cirac, Phys. Rev. A {\bf{81}}, 052338 (2010).


\bibitem{Zgid} D. Zgid, E. Gull
and G. K.-L. Chan, Phys. Rev. B {\bf{86}}, 165128 (2012).



\bibitem{DMFT-1} A. Georges, G. Kotliar, W. Krauth and M. J. Rozenberg, Rev. Mod. Phys. {\bf{68}}, 13 (1996).


\bibitem{maier2005} T. Maier, M. Jarrell, T. Pruschke and
    M. H. Hettler, Rev. Mod. Phys. 77, 1027 (2005).


\bibitem{stanescu2006} T. D. Stanescu, M. Civelli, K. Haule and
    G. Kotliar, Ann. Phys. 321, 1682 (2006).

\bibitem{moukouri2001}
S. Moukouri and M. Jarrell, Phys. Rev. Lett. {\bf{87}}, 167010 (2001).


\bibitem{huscroft2001}
C. Huscroft, M. Jarrell, Th. Maier, S. Moukouri and A. N. Tahvildarzadeh, Phys. Rev. Lett. {\bf{86}}, 139 (2001).



\bibitem{aryanpour2003}
K. Aryanpour, M. H. Hettler and M. Jarrell, Phys. Rev. B {\bf{67}}, 085101 (2003).


\bibitem{DVP-2} M. Potthoff, Eur. Phys. J. B {\bf{32}}, 429 (2003).



\bibitem{Knizia-Chan} G.  Knizia and G. K.-L. Chan, Phys. Rev. Lett. {\bf{109}}, 186404 (2012).

\bibitem{Irek-DMET} I. W. Bulik, G. E. Scuseria and J. Dukelsky, Phys. Rev. B {\bf{89}}, 035140 (2014).

\bibitem{LIEB} E. H. Lieb and F. Y. Wu, Phys. Rev. Lett. {\bf{20}}, 1445 (1968).

\bibitem{BETHE} H. Bethe, Z. Phys. {\bf{71}}, 205 (1931).


\bibitem{Tomita-3} N. Tomita, Phys. Rev. B {\bf{79}}, 075113 (2009).

\bibitem{Tomita-1} N. Tomita, Phys. Rev. B {\bf{69}}, 045110 (2004).

\bibitem{Tomita-2} N. Tomita and S. Watanabe, Phys. Rev. Lett. {\bf{103}}, 116401 (2009).

\bibitem{Tomita-2011PRB} F. Satoh, M. A. Ozaki, T. Maruyama and N. Tomita, Phys. Rev. B {\bf{84}}, 245101 (2011).


\bibitem{Ogata-Shiba} M. Ogata and H. Shiba, Phys. Rev. B {\bf{41}}, 2326 (1990).

\bibitem{Voit} J. Voit, Phys. Rev. B {\bf{47}}, 6740 (1993).


\bibitem{Kim} B. J. Kim, H. Koh, E. Rotenberg, S. -J. Oh, H. Eisaki, N. Motoyama, S. Uchida, T. Tohoyama,
S. Maekawa, Z. X. Shen and C. Kim, Nature {\bf{2}}, 397 (2006).


\bibitem{KMShen} K. M. Shen, F. Ronning, D. H. Lu, W. S. Lee, N. J. C. Ingle, W. Meevasana, F. Baumberger,
A. D. Damascelli, N. P. Armitage, L. L. Miller, Y. Kohsaka, M. Azuma, M. Takano, H. Takagi and
Z. X. Shen, Phys. Rev. Lett. {\bf{93}}, 267002 (2004).





\bibitem{Carlos-Hubbard-1D} K.W. Schmid, T. Dahm, J. Margueron
 and H. M\"uther, Phys. Rev. B {\bf{72}}, 085116 (2005).

\bibitem{Rayner-2D-Hubbard-PRB-2012} R. Rodr\'iguez-Guzm\'an, K.W. Schmid, C. A. Jim\'enez-Hoyos and G. E. Scuseria, Phys. Rev. B {\bf{85}}, 245130 (2012).


\bibitem{non-unitary-paper-Carlos}
 C. A. Jim\'enez-Hoyos, R. Rodr\'iguez-Guzm\'an and
G. E. Scuseria, Phys. Rev. A {\bf{86}}, 052102 (2012).


\bibitem{rayner-Hubbard-1D-FED2013} R. Rodr\'iguez-Guzm\'an, C. A. Jim\'enez-Hoyos, R. Schutski
and G. E. Scuseria, Phys. Rev. B {\bf{87}}, 235129 (2013).

  


\bibitem{rs} P. Ring and P. Schuck, {\em The Nuclear Many-Body Problem} (Springer,
Berlin, 1980).

\bibitem{Carlo-review} K. W. Schmid, Prog. Part. Nucl. Phys. {\bf{52}}, 565 (2004).

\bibitem{rayner-GCM-paper} R. Rodr\'iguez-Guzm\'an, J. L. Egido and L. M. Robledo, Nucl. Phys. A {\bf{709}}, 201 (2002)

\bibitem{rayner-GCM-parity} R. Rodr\'iguez-Guzm\'an, L. M. Robledo and P. Sarriguren, Phys. Rev. C {\bf{86}}, 034336  (2012).


\bibitem{Carlos-Rayner-Gustavo-VAMPIR-molecules} C. A. Jim\'enez-Hoyos, R. Rodr\'iguez-Guzm\'an and
G. E. Scuseria, J. Chem. Phys. {\bf{ 139}}, 224110 (2013).


\bibitem{Carlos-Rayner-Gustavo-FED-molecules} C. A. Jim\'enez-Hoyos, R. Rodr\'iguez-Guzm\'an and
G. E. Scuseria, J. Chem. Phys. {\bf{ 139}}, 204102 (2013).



\bibitem{PQT-reference-1} G. E. Scuseria, C. A. Jim\'enez-Hoyos, T. M. Henderson, K. Samanta
and J. K. Ellis, J. Chem. Phys. {\bf{135}}, 124108 (2011)

\bibitem{PQT-reference-2} C. A. Jim\'enez-Hoyos, T. M. Henderson, T. Tsuchimochi
and G. E. Scuseria, J. Chem. Phys. {\bf{136}}, 164109 (2012)

\bibitem{PQT-reference-3} K. Samanta, C. A. Jim\'enez-Hoyos
and G. E. Scuseria, J. Chem. Theory Comput. {\bf{8}}, 4944 (2012).





\bibitem{Blaizot-Ripka}  J.-P. Blaizot and G. Ripka, {\em Quantum Theory of Finite Fermi Systems}
(The MIT Press, Cambridge, MA, 1985).


\bibitem{Fukutome-original-RSHF} H. Fukutome, Prog. Theor. Phys. {\bf{80}}, 417 (1988); {\bf{81}}, 342 (1989).



\bibitem{Yamamoto-1} S. Yamamoto, A. Takahashi and H. Fukutome, J. Phys. Soc. Jpn.
{\bf{60}}, 3433 (1991).

\bibitem{Yamamoto-2} S. Yamamoto and H. Fukutome, J. Phys. Soc. Jpn.
{\bf{61}}, 3209 (1992).

\bibitem{Ikawa-1993} A. Ikawa, S. Yamamoto, and H. Fukutome, J. Phys. Soc. Jpn.
{\bf{62}}, 1653 (1993).

\bibitem{Gutzwiller_method} M.C. Gutzwiller, Phys. Rev. Lett. {\bf{10}}, 159 (1963).




\bibitem{Laimis-paper} L. Bytautas, Carlos A. Jim\'enez-Hoyos, R. Rodr\'{\i}guez-Guzm\'an
and Gustavo E. Scuseria,  Mol. Phys., {\it{in press}}.


\bibitem{Perlemov} A.M. Perlemov, Sov. Phys. Usp. {\bf{20}}, 703 (1977).


\bibitem{PaperwithShiwei} H. Shi, C. A. Jim\'enez-Hoyos, R. Rodr\'iguez-Guzm\'an, G. E. Scuseria
and S. Zhang, arXiv//cond-mat.str-el//1402.0018 (2013).


\bibitem{CPMC-ref} S. Zhang, J. Carlson and J. E. Gubernatis, Phys. Rev. B {\bf{55}}, 7464 (1997).


\bibitem{ALPS} A. F. Albuquerque et al., J. Magn. Magn. Mater.  {\bf{310}}, 1187
(2007); B. Bauer et al., J. Stat. Mech.  (2011) P05001.






\bibitem{StuberPaldus} J. L. Stuber and J.Paldus, {\em
Symmetry Breaking in the Independent Particle
Model}. Fundamental World of Quantum Chemistry: A Tribute Volume to
the Memory of Per-Olov L\"owdin; Edited by E. J. Brandas
and E. S Kryachko (Kluwer Academic
Publishers: Dordrecht, The Netherlands, 2003).

\bibitem{HFclassification} H. Fukutome, Int. J. Quantum Chem., {\bf{20}},  955 (1981).



\bibitem{Edmonds} A. R. Edmonds, {\em Angular Momentum in Quantum Mechanics}, Princenton
Univ. Press, Princenton (1957).

\bibitem{Ashcroft-Mermin-book} N. W. Ashcroft and N.D. Mermin, {\it{Solid State Physics}}, Saunders
College, 1976.


\bibitem{Schmid-Gruemmer-1984} K. W. Schmid, F. Gr\"ummer and A. Faessler, Phys. Rev. C {\bf{29}}, 291 (1984).



\bibitem{quasi-Newton} D. C. Liu and J. Nocedal, Math. Program. B {\bf{45}}, 503 (1989).




\bibitem{CrossOverSS} M. Imada, N. Furukawa and T. M. Rice, J. Phys. Soc. Jpn.
{\bf{61}}, 3861 (1992).


\bibitem{Horovitz} B. Horovitz, in {\em Solitons}, edited by S. E. Trullinger , V. E. Zakharov
and V. L. Pokrovsky (Elsevier, Amsterdam, 1986).

\bibitem{Sorella-1990} S. Sorella, A. Parola, M. Parrinello
and E. Tosati, Europhys. Lett. {\bf{12}}, 721 (1990).

\bibitem{Qin-S2kf} S. Qin, S. Liang, Z. Su and L. Yu, Phys. Rev. B {\bf{52}}, 5475 (1995).

\bibitem{Solyom} J. S\'olyom, Adv. Phys. {\bf{28}}, 201 (1979).

\bibitem{SchulzPRL1990} H. J. Schulz, Phys. Rev. Lett. {\bf{64}}, 2831 (1990).


\bibitem{Hirsch-S2kf} J. E. Hirsch and D. J. Scalapino, Phys. Rev. B {\bf{27}}, 7169 (1983).



\bibitem{Imada-S2kf} M. Imada and Y. Hatsugai, J. Phys. Soc. Jpn.
{\bf{58}}, 3752 (1989).

\bibitem{AraGo} A. Go and G. S. Jeon, J. Phys.: Condens. Matter {\bf{21}}, 485602 (2009).




\bibitem{Senechal-scs} D. S\'en\'echal, D. Perez and M. Pioro-Ladri\'ere, Phys. Rev. Lett. {\bf{84}}, 522 (2000).


\bibitem{PENcPRL1996} K. Penc, K. Hallberg, F. Mila
and H. Shiba, Phys. Rev. Lett. {\bf{77}}, 1390 (1996);
J. Favand, S. Haas, K. Penc, F. Mila
and E. Dagotto, Phys. Rev. B {\bf{55}}, R4859 (1997); A. Parola
and S. Sorella, Phys. Rev. B {\bf{45}}, R13156 (1992); M. Ogata, T. Sugiyama and
 H. Shiba, Phys. Rev. B {\bf{43}}, 8401 (1991); M. Ogata and  H. Shiba, Phys. Rev. B {\bf{41}}, 2326 (1990).

\bibitem{Yokoyamma-Shiba} H. Yokoyama and H. Shiba, J. Phys. Soc. Jpn. {\bf{56}}, 3582 (1987).


\bibitem{Scmid-RPA} K. W. Schmid, M. Kyotoku, F. Gr\"ummer and A. Faessler, Ann. Phys.  {\bf{190}}, 182 (1989).

\bibitem{Nishiyama-RPA} S. Nishiyama, Prog. Theor. Phys. {\bf{69}}, 100 (1983).

\bibitem{optical-Dagotoo} A. Moreo and E. Dagotto, Phys. Rev. B {\bf{42}}, 4786 (1990).

\bibitem{optical-Scalapino} R. M. Fye, M. J. Martins, D. J. Scalapino, J. Wagner and W. Hanke, Phys. Rev. B {\bf{45}}, 7311 (1992).

\bibitem{Jeckelmann} E. Jeckelmann, F. Gebhard and F. H. L. Essler, Phys. Rev. Lett. {\bf{85}}, 3910 (2000).


\end{thebibliography}
\end{document}